\documentclass[a4paper, 10pt, oneside, number,sort&compress, 3p, twocolumn]{elsarticle}

\journal{}

\usepackage{graphicx}
\usepackage{dcolumn}
\usepackage{bm}

\usepackage[english]{babel}
\usepackage[latin1]{inputenc} 

\usepackage{epstopdf}
\usepackage{amsfonts}
\usepackage{color}
\usepackage{epsfig}
\usepackage{mathrsfs}
\usepackage{amsmath,amssymb}
\usepackage{subfigure}

\usepackage{framed}
\usepackage{lipsum}
\usepackage[svgnames]{xcolor}
\definecolor{shadecolor}{named}{LightGrey}

\usepackage{float}
\floatstyle{plaintop}
\restylefloat{table}
\usepackage[tableposition=top]{caption}

\usepackage{blindtext}

\usepackage{fancyhdr}
\usepackage{multicol}

\usepackage{pifont}%
\usepackage{geometry}%
\usepackage{fleqn}
\usepackage{txfonts}
\usepackage{hyperref}

\usepackage{lineno}

\def\Xmax{\ifmmode {X_\mathrm{max}}\else
                   {$X_\mathrm{max}$}\fi\xspace}%

\begin{document}

\begin{frontmatter}

\title{Muon tomography imaging algorithms for nuclear threat detection inside large volume containers with the \textit{Muon Portal} detector}

\author[INAF]{S. Riggi\corref{cor}}
\ead{simone.riggi@ct.infn.it}\cortext[cor]{Corresponding author.}
\author[INAF]{V. Antonuccio-Delogu}
\author[INAF]{M. Bandieramonte}
\author[INAF]{U. Becciani}
\author[INAF]{A. Costa}
\author[UNICT,INFN]{P. La Rocca}
\author[INAF]{P. Massimino}
\author[UNICT,INFN]{C. Petta}
\author[INAF]{C. Pistagna}
\author[UNICT,INFN]{F. Riggi}
\author[INAF]{E. Sciacca}
\author[INAF]{F. Vitello}

\address[INAF]{INAF - Osservatorio Astrofisico di Catania, Italy}
\address[UNICT]{Dip. di Fisica e Astronomia, Universit\`{a} di Catania, Italy}
\address[INFN]{INFN Section of Catania, Italy}

\begin{abstract}
Muon tomographic visualization techniques try to reconstruct a 3D image as
close as possible to the real localization of the objects being probed.
Statistical algorithms under test for the reconstruction of muon
tomographic images in the Muon Portal Project are here discussed.
Autocorrelation analysis and clustering algorithms have been employed
within the context of methods based on the Point Of Closest Approach
(POCA) reconstruction tool. An iterative method based on the
log-likelihood approach was also implemented. Relative merits of all such
methods are discussed, with reference to full \textsc{Geant}4 simulations of
different scenarios, incorporating medium and high-Z objects inside a
container.
\end{abstract}


\end{frontmatter}

\section{Introduction}\label{IntroductionSection}
Muon tomography is a technique employing the scattering of secondary cosmic muons inside a material, to reconstruct a 3D image as close as possible to the true 
localization of the objects inside the volume to be inspected. Due to the dependence of the scattering angle on the atomic number $Z$ of the material, this
technique is particularly promising to search for the presence of high-$Z$ materials inside large volumes - such as containers - even in presence 
of additional, low- and medium-$Z$, objects.\\To reach a good precision in the reconstruction of the tomographic image a good tracking muon detector
is required, able to reconstruct on a event-by-event basis the track of the muon before and after traversing the volume, even in presence of multiple hits
generated by the structure itself (mechanical structure, empty container, \dots) or by the surrounding materials (roof, buildings, soil, \dots).
Spatial and angular resolutions are then mandatory in this respect, to provide a good description of the incoming and outgoing muon tracks. Such performances
may be achieved with several detection technologies, some of which have inherent good spatial resolution (for instance multiwire gas detectors) whereas others
with worse intrinsic resolution (such as segmented scintillator strips) may reach good results depending on the number of detection planes and 
relative distance between them. On the other side, imaging algorithms of good quality and able to produce results in a small CPU time are an essential tool
for the reconstruction of tomographic images. Relatively short computing times, few minutes at most, are infact required at the inspection site, 
in order to follow the real container flux without significantly interfering with the usual harbour activities.  
A large effort is at present pursued by the different groups working in the field to set up appropriate 
software tools for such task, which involve not only a proper reconstruction, in nearly real-time, of the image, but also good rendering techniques to provide the user
with an easy-to-interpret tomographic image.\\In the framework of a new Project aiming at the construction of a real scale prototype of scanning muon detector
for containers, an important part of our efforts has been concentrated on testing both traditional and new numerical techniques devoted to this task.\\In the 
present paper we report a comparison of several methods which, starting from raw data, produce a tomographic image of the objects contained in a large volume.
Data for such comparison were provided by detailed \textsc{Geant4} simulations incorporating the knowledge of the mechanical structure of the detector and 
of the container under inspection and the physical interactions of the muons (with realistic energy and angular distributions) with all materials. In addition
to the simplest method, based on the reconstruction of the Point-Of-Closest-Approach, also methods based on autocorrelation analysis, clustering and log-likelihood
algorithms have been tested under different scenarios.\\Section \ref{DetectorSection} reports a brief description of the detector prototype under construction,
while Section \ref{AlgorithmSection} discusses the main ingredients of all such methods. Section \ref{ResultsSection} reports the application of these algorithms 
to different sets of simulated data. Finally, Section \ref{SummarySection} is devoted to a discussion of the main aspects which could be improved in the future,
in order to arrive to a real-time tomographic analysis for such detector.

\section{Overview of the \textit{Muon Portal} project} \label{DetectorSection}
The Muon Portal project \cite{RiggiECRS,LoPresti,MuonPortalWebsite} is a prototype of a dedicated particle detector for the inspection of harbour containers through the 
technique of muon tomography. 
The experimental setup is based on four XY detector planes, each providing the X and Y position measurements,
two placed below and two above the volume to be inspected. The size of each plane is optimized to fit that of a real TEU (Twenty-foot Equivalent Units)
container, namely 5.9 m $\times$ 2.4 m $\times$ 2.4 m.\\To favour the detector assembly and its maintenance, each plane is divided into 6 modules 
of size 1 m $\times$ 3 m arranged to cover the above specified detector area with minimal dead surfaces. 
Each module is hosted inside a dedicated casing providing the mechanical 
support for the detector planes. The mechanical structure is designed to minimize the amount of material traversed by the cosmic ray muons.
A dummy mechanical structure is also being designed to be inserted between the intermediate detector planes to emulate a real container volume.\\
Each module is segmented into 100 strips of extruded plastic scintillators (1 $\times$ 1 $\times$ 300 cm$^3$) with two embedded wavelength-shifting (WLS) fibres
to collect the light produced inside the scintillator bars. Each fiber is coupled at one end to Silicon Photomultipliers (SiPMs), designed ad-hoc for 
the project to maximize the light yield with reasonable cost requirements.\\
More details concerning the detector geometry, electronic readout and channel reduction mechanism can be found in \cite{RiggiECRS, LoPresti}.\\
The expected detector acceptance $\mathcal{A}$ to a flux of cosmic ray muons has been evaluated from detailed detector simulations, yielding 
$\mathcal{A}$=10 m$^{2}$sr, corresponding to a number of expected events of 
$\sim$2$\times$10$^{5}$ for a scanning time of $\Delta$t=5 minutes and a standard cosmic ray flux $\phi_{\mu}$= 1 m$^{-2}$s$^{-1}$ integrated over the solid angle.
\\The angular accuracy for muon track 
reconstruction has been found of the order of 0.25$^{\circ}$ from toy simulations in which only the impact of the position resolution is 
considered. It increases at $\sim$0.5$^{\circ}$ in detailed \textsc{Geant4} simulations including also multiple scattering effects.\\The precision 
on the determination of the scattering data, scattering angle and lateral displacement, which are relevant for tomography imaging studies, 
needs to be estimated too. The analysis yields a scattering angle uncertainty of $\sim$0.7$^{\circ}$ and a lateral
displacement uncertainty of the order of 2 cm. This imposes a limit on the minimum size of the threat objects that can be identified with reasonable accuracy
inside the container volume and within reasonable scanning times, typically not smaller than 5 cm.

\section{Statistical methods for muon tomography imaging}\label{AlgorithmSection}
In this section we briefly report details on the statistical algorithms adopted for tomographic image reconstruction.
They have been developed in C++ using links to \textsc{Geant4} \cite{GEANT4} and \textsc{Root} \cite{ROOT} frameworks for 
detector geometry building and navigation and mathematical routines.

\subsection{\textsc{POCA}-based methods}
\begin{figure}[!h]
\centering
\subtable[POCA sketch]{\includegraphics[scale=0.22]{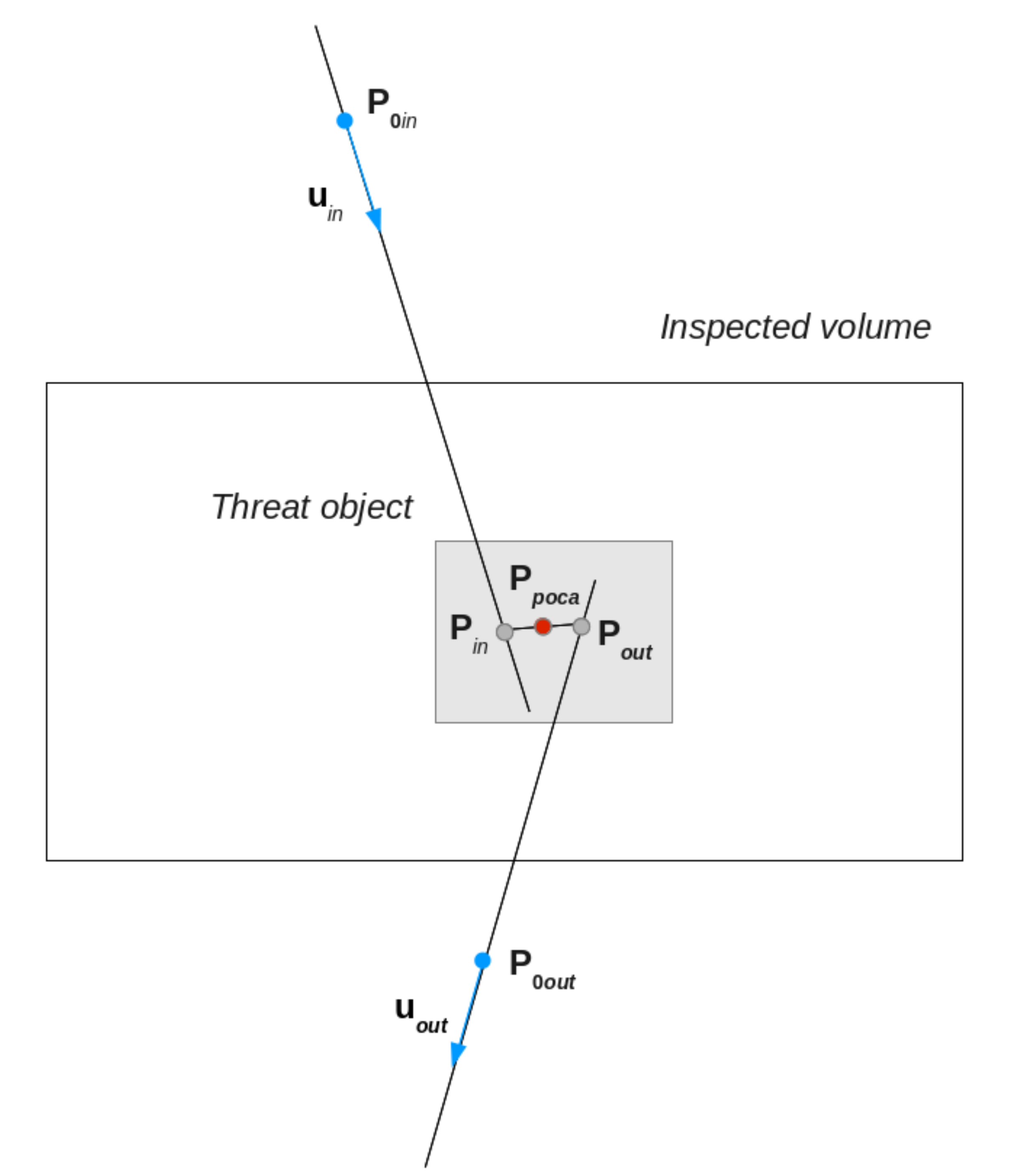}}%
\vspace{0.1cm}
\caption{A sketch illustrating the working principle of the POCA method.}
\label{POCASketchFig}
\end{figure}

The simplest algorithm in the field is the \textit{Point Of Closest Approach} (\textsc{Poca}) which makes the simplified assumption that the muon scattering 
occurs in a single-point.
It therefore searches for the geometrical point of closest approach $\mathbf{P}_{poca}$=$\frac{1}{2}(\mathbf{P}_{in}+\mathbf{P}_{out})$
between the incoming $\mathbf{u}_{in}$ and outcoming $\mathbf{u}_{out}$ reconstructed track directions with respect to the inspected volume 
(see sketch in Figure \ref{POCASketchFig}):
\begin{eqnarray}
\mathbf{P}_{in,out}&=& \mathbf{P_{_{0}}}_{in,out}+t_{in,out}\mathbf{u}_{in,out}\\
t_{in}&=& (b e-c d)/\Delta\\
t_{out}&=& (a e-b d)/\Delta
\end{eqnarray} 
where $\mathbf{P_{_{0}}}_{in,out}$ are two points on the incoming and outgoing tracks, 
$a=\mathbf{u}_{in}\cdot\mathbf{u}_{in}$, $b=\mathbf{u}_{in}\cdot\mathbf{u}_{out}$, $c=\mathbf{u}_{out}\cdot\mathbf{u}_{out}$, 
$d=\mathbf{u}_{in}\cdot\mathbf{w}$, $e=\mathbf{u}_{out}\cdot\mathbf{w}$, $\Delta=ac-b^{2}$, $w=\mathbf{P_{_{0}}}_{in}-\mathbf{P_{_{0}}}_{out}$.

Such method is of easy implementation and provides fast results, useful as first-order approximation
to the problem or as a starting approximation for more detailed algorithms. However, it neglects the multiple scatterings throught the volume material and 
therefore has the drawback of providing poor-resolution images, it is quite sensitive to the presence of shield materials located above or below 
the potential threat and cannot localize very well materials at the volume borders.
This motivates the implementation of the log-likelihood algorithm discussed below, which is based on more realistic physical and statistical assumptions 
and allows to face the problems encountered with the \textsc{Poca} algorithm. 
It is also desirable to have additional ``grid-free'' statistical methods for tomography analysis, which do not require to assume a predefined grid dividing the 
inspected volume into three-dimensional voxels.
We explored in next paragraphs two alternative methods, based on the POCA observable: the two-points autocorrelation analysis and 
density-based clustering algorithms.

\subsubsection[AutocorrelationSection]{Autocorrelation analysis}\label{AutocorrelationSection}
The two-point autocorrelation function, hereafter denoted \emph{2pt-ACF} for brevity, is one of the main statistics generally used to describe the distribution of 
galaxies and to search for localized excess of data observations at certain scales in a volume with respect to a homogeneous random distribution. 
It is therefore well suited also for the problem of tomography imaging where we need to search for a density excess of POCA observations inside the container with 
respect to a normal situation, for instance an empty container.\\
Following Peebles \cite{Peebles} the 2pt-ACF $\xi(r)$ defines the probability $dP$ to find simultaneously two objects at a distance $r$ from each other 
within two volume elements $dV_{1}$ and $dV_{2}$ in a data sample with event density $n$:
\begin{equation}
dP = n^{2}[1 + \xi(r)]dV_{1}dV_{2}
\end{equation}
A positive correlation ($\xi>$0) at distance $r$ indicate clustering at such scale, anticorrelation ($\xi<$0) indicate that the objects tend to avoid each other, while
$\xi\sim$0 is relative to an homogeneous distribution without significative clusters.\\
For practical purposes the \emph{2pt-ACF} can be computed from a sample of objects counting the pairs of observations at different separations $r$. Four estimators
are generally used in literature: Peebles-Hauser $\hat{\xi}_{PH}$ \cite{PeeblesHauser}, Davis-Peebles $\hat{\xi}_{DP}$ \cite{DavisPeebles}, Hamilton $\hat{\xi}_{H}$
\cite{Hamilton} and Landy-Szalay $\hat{\xi}_{LS}$ \cite{LandySzalay}. They require to calculate the number of pairs $DD(r)$, $RR(r)$, $DR(r)$ at distance $r$ respectively present in the data set (data-data), 
in a random data set (random-random) and in the data-random set (data-random):
\begin{eqnarray}
\hat{\xi}_{PH}(r)&=& \frac{N_{RR}}{N_{DD}}\frac{DD(r)}{RR(r)}-1\\%
\hat{\xi}_{DP}(r)&=& \frac{N_{DR}}{N_{DD}}\frac{DD(r)}{DR(r)}-1\\%
\hat{\xi}_{H}(r)&=& \frac{N_{DR}^{2}}{N_{DD}N_{RR}}\frac{DD(r)RR(r)}{[DR(r)]^{2}}-1\\%
\hat{\xi}_{LS}(r)&=& 1+\frac{N_{RR}}{N_{DD}}\frac{DD(r)}{RR(r)}-2\frac{N_{RR}}{N_{DR}}\frac{DR(r)}{RR(r)}%
\end{eqnarray}
with $N_{D}$, $N_{R}$ total number of observations present in the data and random data sets and where $N_{DD}= N_{D}(N_{D}-1)/2$, $N_{RR}= N_{R}(N_{R}-1)/2$ 
and $N_{DR}= N_{D}N_{R}$ are the total number of corresponding pairs in the data-data, random-random, data-random catalogues.
Such estimators take into account the edge effect due to the fact that it is not always possible to fit in complete spheres of radius $r$ at every position
within a survey volume, for example at the container borders.\\
The above estimators define spatial correlation only. To seach for both spatial and angular correlations we introduced a weight 
$w_{ij}=\theta_{i}^{\alpha}+\theta_{j}^{\alpha}$ ($\alpha$=2) for each observation pair $ij$ with scattering angles ($\theta_{i}$, $\theta_{j}$) and computed a 
weighted correlation estimator $\xi_{w}(r)$. 

\begin{figure}[!h]
\centering
\subtable[EM-ML sketch]{\includegraphics[scale=0.22]{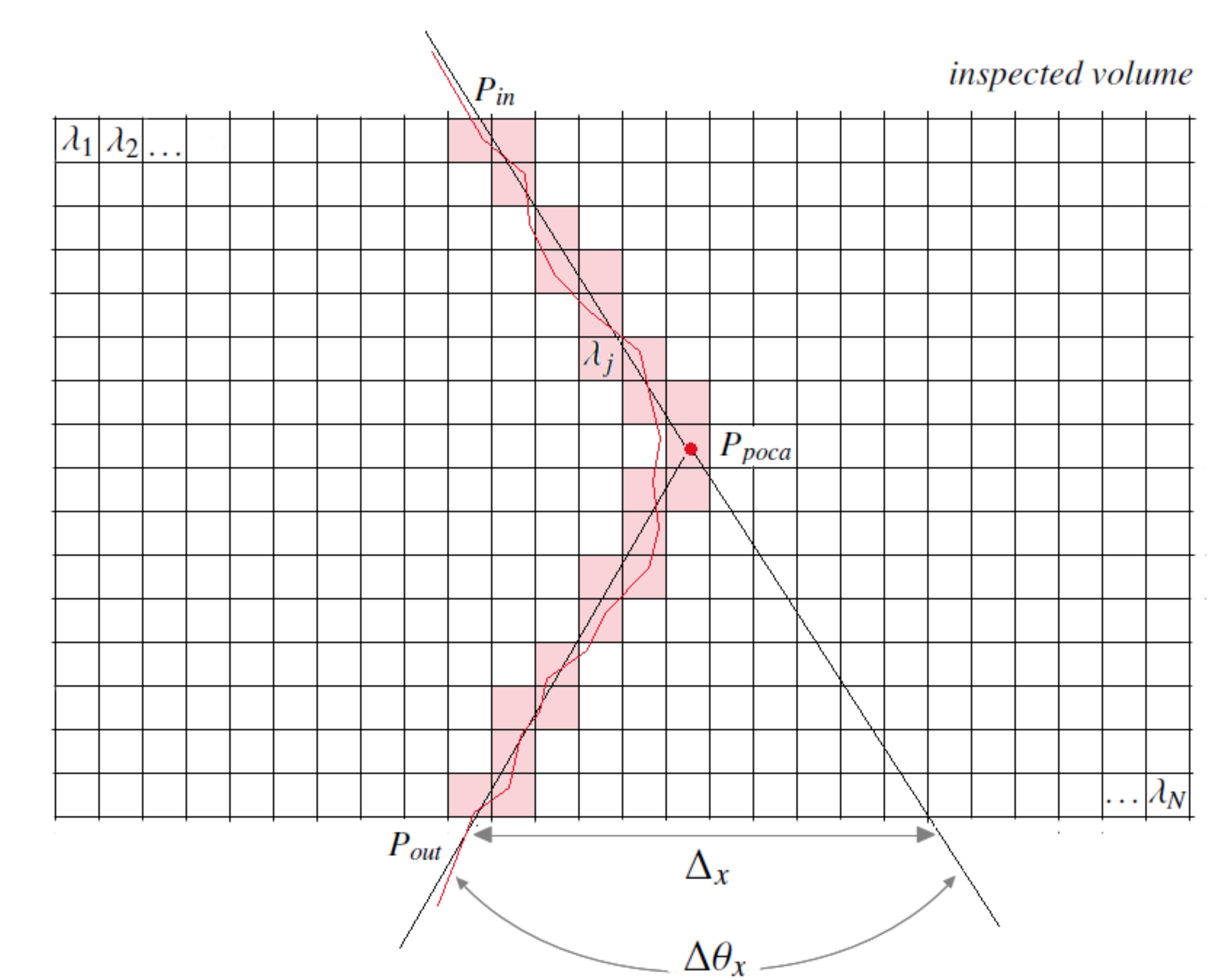}}%
\vspace{0.1cm}
\caption{A sketch illustrating the working principle of the EM-ML method.}
\label{EMLLSketchFig}
\end{figure}

\subsubsection[Clustering algorithms]{Clustering algorithms}\label{ClusteringSection}
High-Z materials are imaged with the POCA method as regions of higher densities of unspecified shape with respect to the background. 
It is therefore a natural choice to employ density-based clustering methods in the tomography reconstruction. 
They connect data points within a certain distance threshold $\epsilon$, satisfying a density criterion defined as the minimum number of 
objects $N_{min}$ within $\epsilon$. A cluster of arbitrary shape, in contrast to other clustering methods, is in this way defined by all density-connected objects
plus all objects within the distance range.\\The most popular clustering algorithms is \textsc{dbscan} \cite{DBSCAN}. It requires 
$\epsilon$ and $N_{min}$ as input parameters and it is based on the following steps:

\begin{enumerate}
 \item Choose an arbitrary unvisited data point as starting point;
 \item Find the neighborhood of this point, e.g. all points within the radius $\epsilon$. If more than $N_{min}$ neighborhoods are found around this point then
 a cluster is started and the point marked as visited, otherwise the point is labelled as noise;
 \item If a point is found to be a part of the cluster then its $\epsilon$ neighborhood is also the part of the cluster and the above procedure from step 2 is 
repeated for all $\epsilon$ neighborhood points. This is repeated until all points in the cluster are determined.
  \item A new unvisited point is retrieved and processed, leading to the discovery of a further cluster or noise.
\item This process continues until all points are marked as visited.
\end{enumerate}
Several clustering algorithms were tested, including \textsc{dbscan}. The friends-of-friends algorithm \cite{fof, fof1}, hereafter denoted as \textsc{fof}, 
has been found to provide the best results. The \textsc{fof} is a percolation algorithm normally used to identify dark matter halos from $N$-body simulations. 
It defines uniquely groups that contain all the particles separated by a distance smaller than a given linking length $l_{link}$.
Once the linking length is defined, the algorithm identifies all pairs of particles which have a mutual distance smaller than the linking one. 
These pairs are designated friends, and clusters are defined as sets of particles that are connected by one or more of the friendly relations, 
so that they are friends of friends. The linking length is related to a parameter $l$ that represents the mean interparticle 
separation in simulations (related to the mean number density $\langle n\rangle$ as $l={\langle n\rangle}^{-1/3}$).
Another parameter in FOF algorithm is the minimum number of particles $N_{min}$, in a cluster. The aim is to reject spurious clusters, 
that is groups of friends who do not form persistent objects in the simulation. Choosing $N_{min}$ sufficiently large allows 
to eliminate spurious clusters. In fact it is much more likely that a spurious cluster (noise) involves a small number of points and not viceversa.

\begin{figure*}[!th]


\begin{minipage}[c]{0.5\textwidth}
\centering
\includegraphics[scale=0.3]{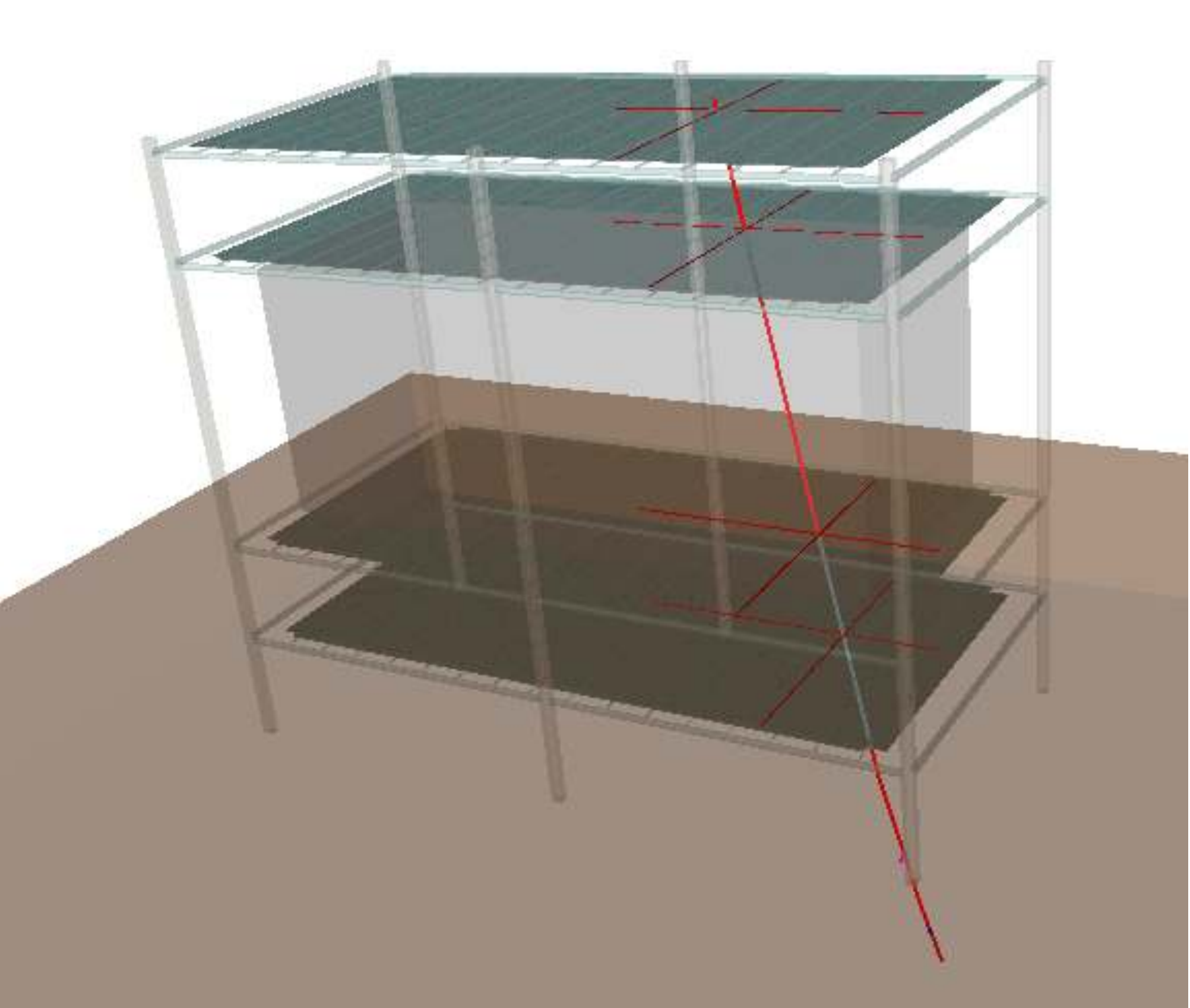}%

\end{minipage}

\vspace{-5.5cm}
\hspace{8cm}
\begin{minipage}[c]{0.6\textwidth}

\includegraphics[scale=0.15]{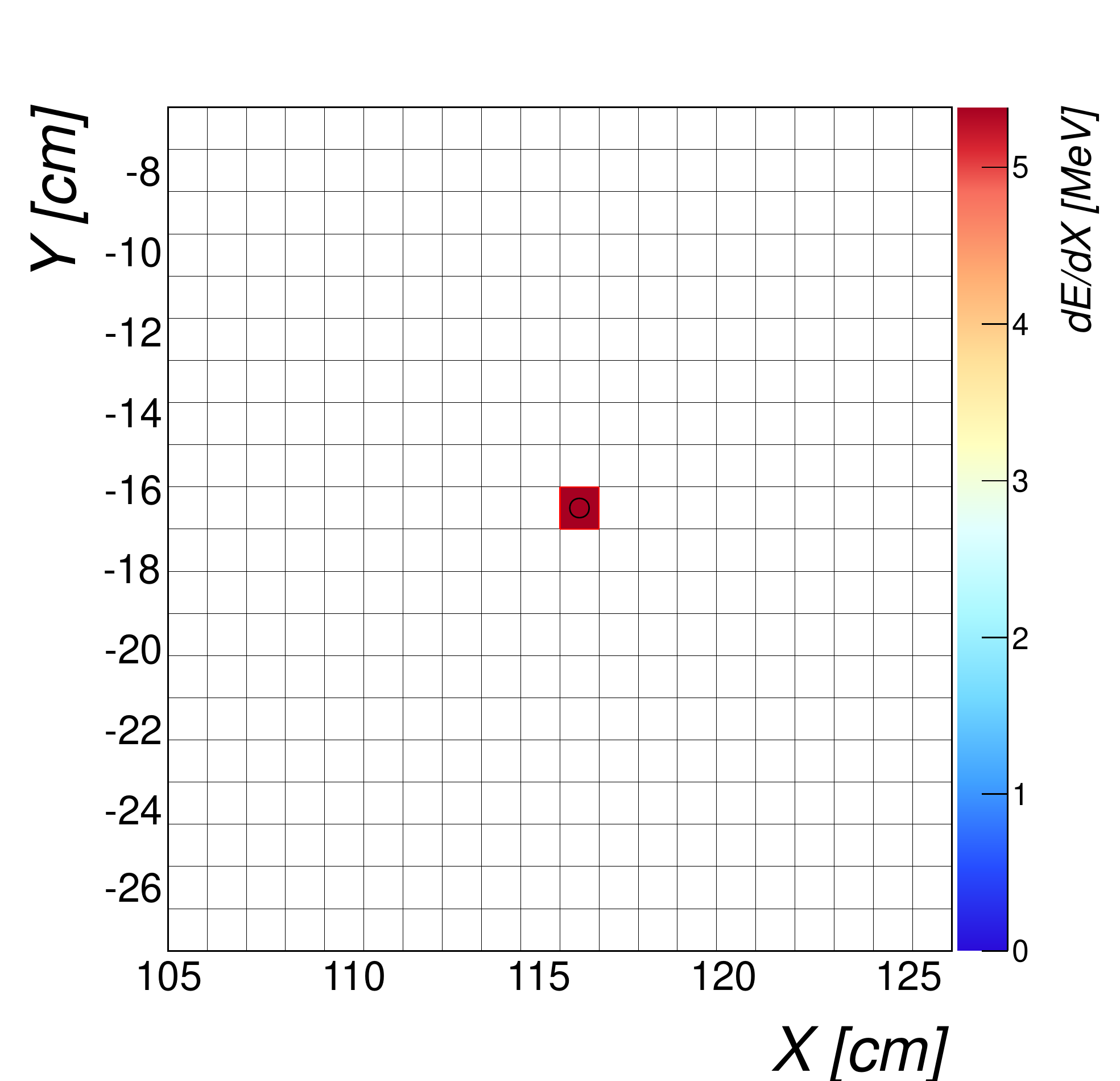}%
\includegraphics[scale=0.15]{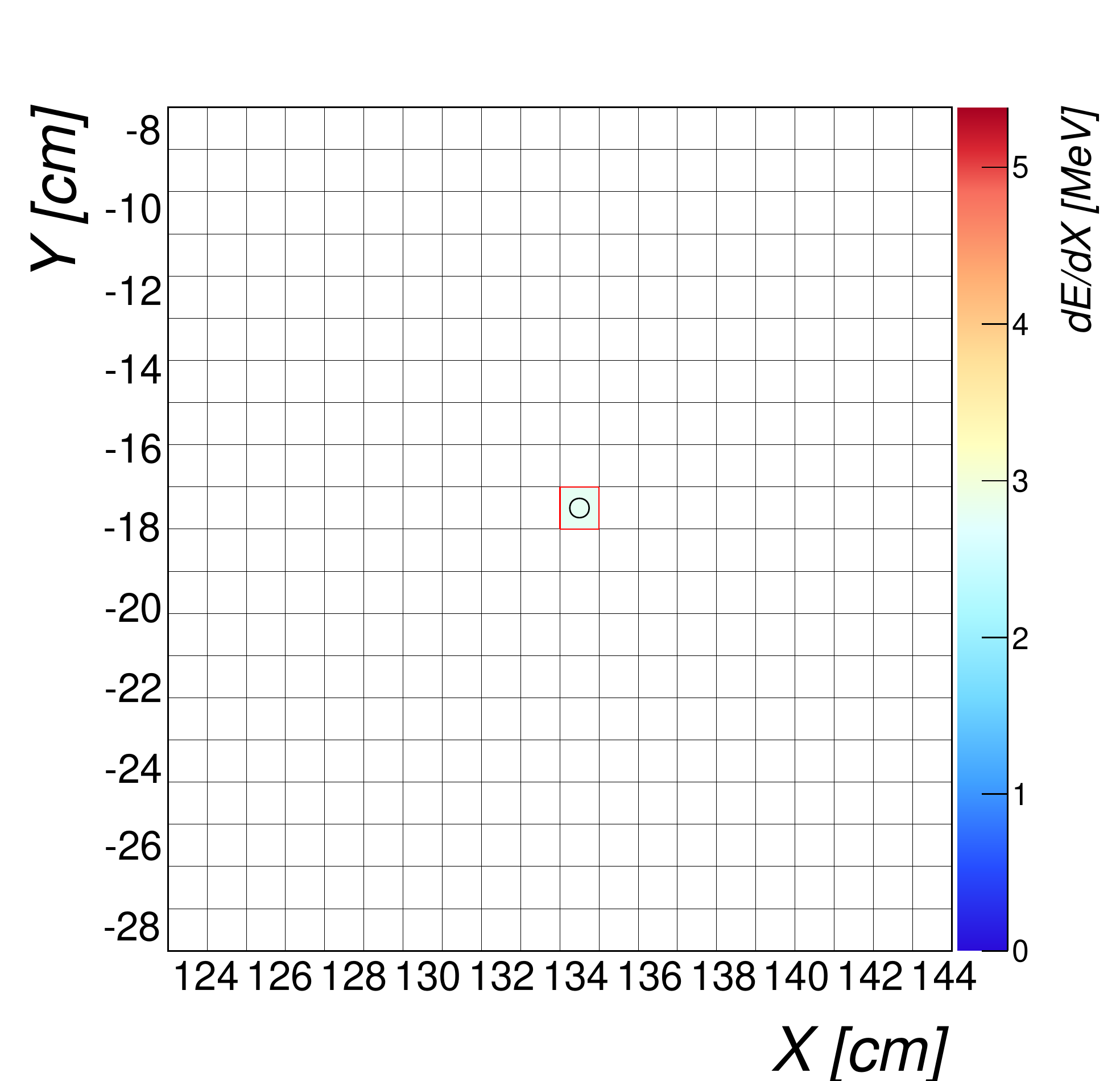}\\%
\includegraphics[scale=0.15]{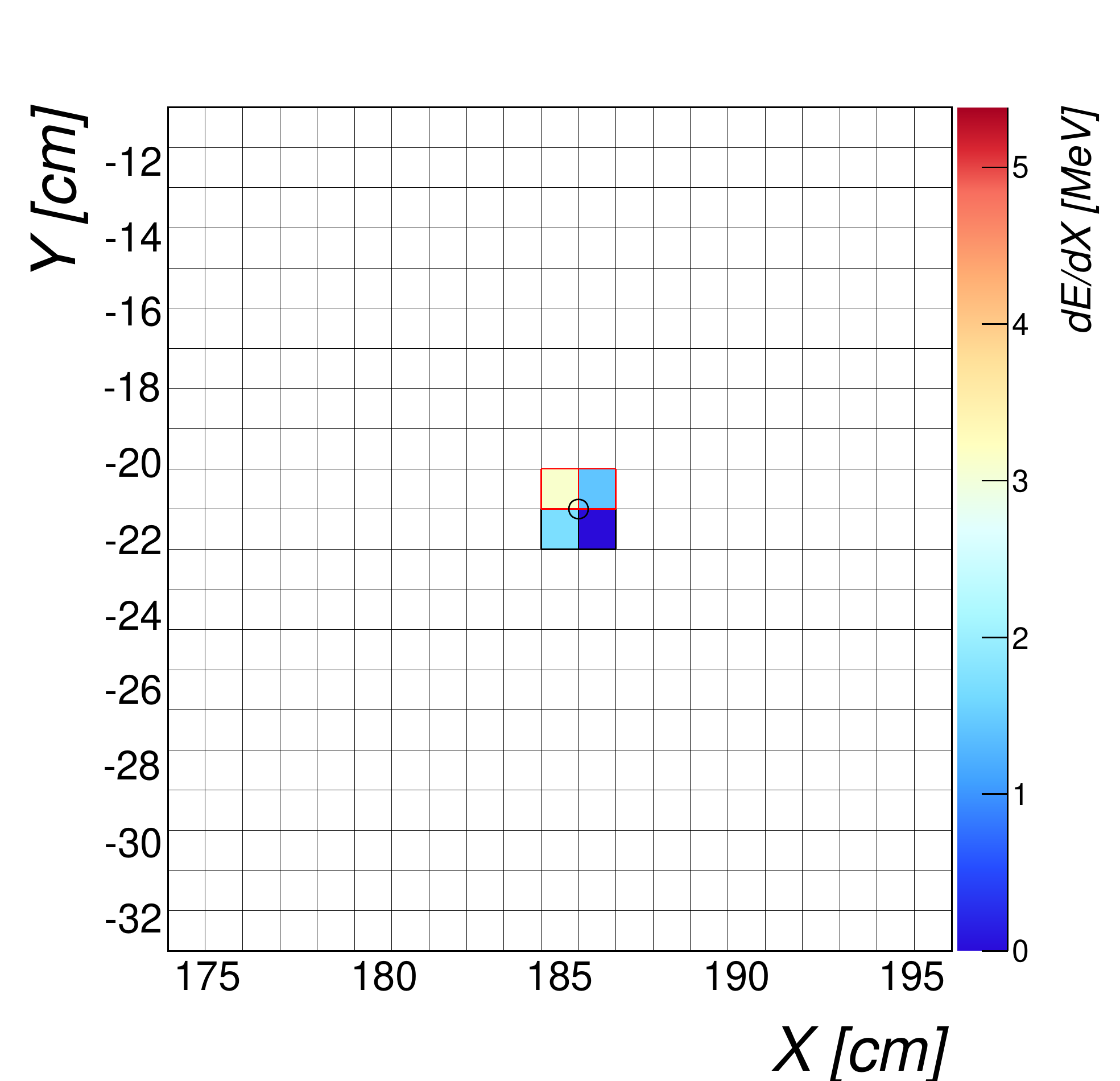}%
\includegraphics[scale=0.15]{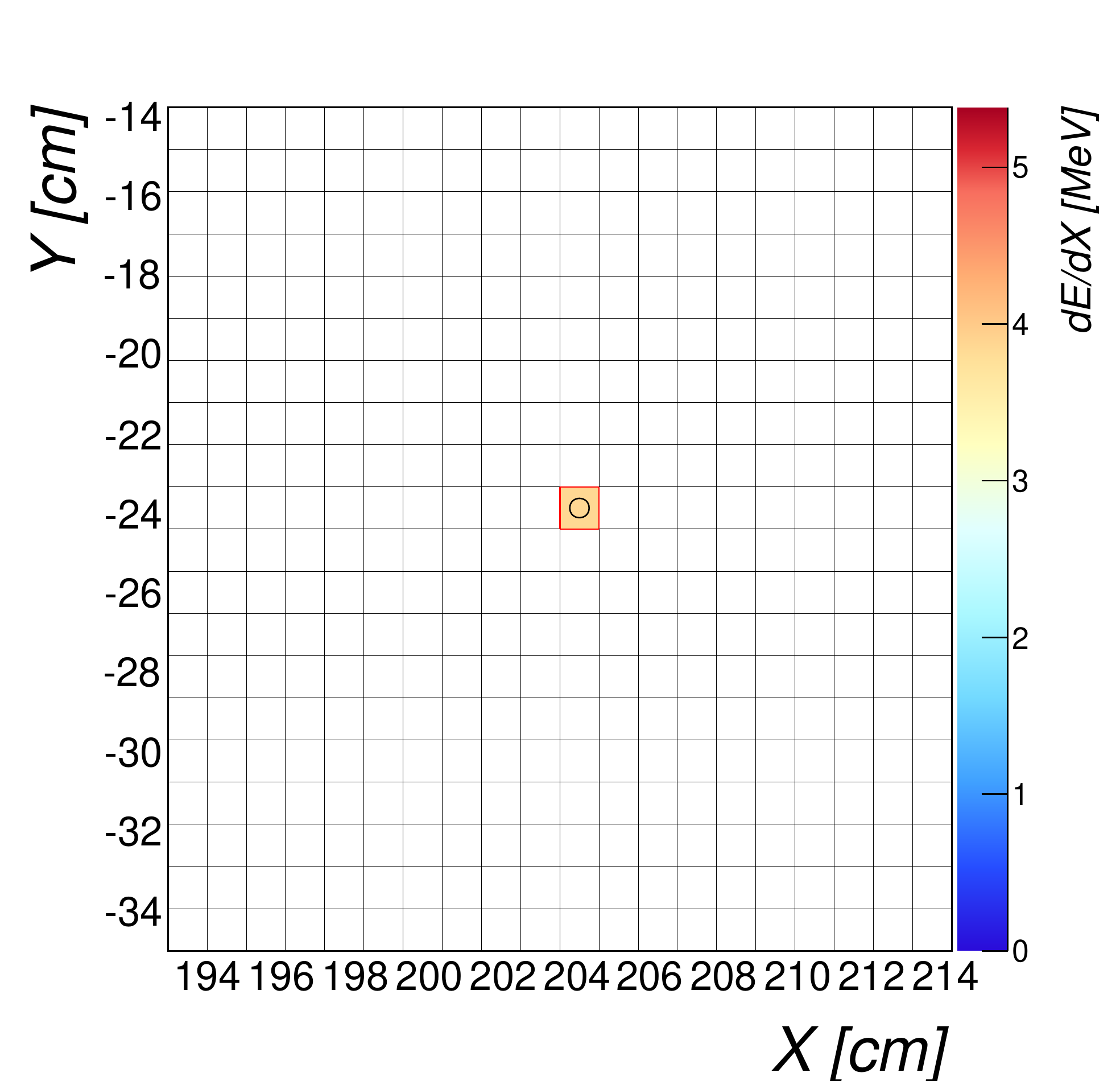}%

\end{minipage}
\vspace{-0.1cm}
\caption{Sample simulated $\mu$ event of energy 1 GeV with hits produced in the four detector planes. The color scale of the plots represents the energy 
deposited in each hit. Hits with red borders are effectively due to the muon while others are spurious.}%
\label{DetectorFig}
\end{figure*}

\subsection[Log-likelihood algorithm]{Maximum likelihood method}
A better statistical treatment of the scattering processes can be done using a log-likelihood approach. It assumes the volume to be imaged divided into $N_{voxels}$ 
three-dimensional voxels or pixels of size $N_{x}\times N_{y}\times N_{z}$. Following the well known Rossi's formula, 
describing the variance of the scattering angle of a particle of momentum $p_{0}$ traversing a material of radiation length $X_{0}$, 
a scattering density $\lambda$ is defined for each voxel and given by:
\begin{equation}\label{ScatteringDensityDefinition}
\lambda(X_{0})= \left(\frac{15\;\mbox{MeV}}{p_{0}}\right)^{2}\frac{1}{X_{0}}
\end{equation}
The determination of $\lambda_{j}$ ($j$=1,\dots $N_{voxels}$) can be done by fitting the scattering data $\mathbf{x}_{i}$= ($\Delta\theta_{x,y}$, $\Delta_{x,y}$) for each $i$-th muon event 
for both x and y coordinates. Such joint distribution for a given scattering layer is with good approximation\footnote{Actually long tails are present 
in the distribution and $\sim$2\% of the data cannot 
be well described by the single gaussian assumption. A gaussian mixture is often used to reproduce the tails.} 
modelled with a bivariate gaussian with covariance matrix $\mathbf{\Sigma}_{i}$ given by:
\begin{equation}
\Sigma_{i}= E_{i}+p_{r,i}^{2}\sum_{j=1}^{N_{voxels}}W_{ij}\lambda_{j} 
\end{equation}
where $E_{i}$ is the measurement error matrix, $W_{ij}$ is the scattering covariance matrix through the $j$ voxel and $p_{r,i}= p/p_{0}$ is the ratio between 
the muon momentum $p$ and the reference momentum $p_{0}$.\\
The log-likelihood $\mathcal{L}$ of a data sample of $N$ muon events is therefore given by:
\begin{equation}
P(\mathbf{x}|\lambda)= \prod_{i=1}^{N}\frac{1}{2\pi|\Sigma_{i}|^{1/2}}\exp\left(-\frac{1}{2}\mathbf{x}_{i}^{T}\mathbf{\Sigma}_{i}\mathbf{x}_{i}\right) 
\end{equation}
\begin{equation}
\mathcal{L}(\mathbf{x}|\lambda)= \frac{1}{2}\sum_{i}^{N}(\log|\Sigma_{i}^{-1}|-y_{i}^{T}\Sigma_{i}^{-1}y_{i}) 
\end{equation}
The scattering densities $\lambda_{j}$ are estimated by maximizing the above log-likelihood. Traditional algorithms, such as those based on Newton-Raphson optimization, 
are limited by the large number of parameters to be determined, i.e. $\sim$5$\times$10$^{4}$ for voxels of size 10 cm, and by considerable computation and storage 
required to compute the Hessian matrix. Schultz et al \cite{Schultz} provided a closed form solution to the problem in the EM formulation, leading to
the following iterative estimation for $\lambda_{j}$:
\begin{align}\label{EMIterationFormula}
\lambda_{j}^{(k+1)}&= \frac{1}{M_{j}}\sum_{i}S_{ij}^{(k)}\\
S_{ij}&= 2\lambda_{j}^{(k)}+p_{r,i}^{2}(\lambda_{j}^{(k)})^{2}(y_{i}^{T}\Sigma_{i}^{-1}W_{ij}\Sigma_{i}^{-1}y_{i}-\mbox{Tr}(\Sigma_{i}^{-1}W_{ij}))
\end{align}
where $M_{j}$ is the number of events traversing voxel $j$. A formula to compute $W_{ij}$ is also available in \cite{Schultz}. In the following we will therefore denote
this method as \emph{EM-ML} for brevity.
\\The algorithm requires the following stages:
\begin{itemize}
 \item Init
 \begin{enumerate}
 \item Reconstruct the scattering data ($\Delta\theta_{x,y}$, $\Delta_{x,y}$)$_{i}$ for each event;
 \item Compute the weight matrices $W_{ij}$ for each event $i$ on the basis of the muon path length through the $j$ voxel. The latter can be estimated 
 assuming a straight line connecting entrance and exit points from the inspected volume, eventually passing from the POCA point, if this is available or trustable. 
 Figure \ref{EMLLSketchFig} shows a sketch of the algorithm raytracing principle.
 The path length calculation is achieved with a standalone \textsc{Geant4} navigator allowing a fast navigation through the container voxelized geometry;
\end{enumerate}
  \item Imaging
  \begin{enumerate}
   \item Assume an initial estimate $\lambda_{j}^{0}$ for $\lambda_{j}$;
   \item Iterate formula (\ref{EMIterationFormula}) until convergence or early stopping;
  \end{enumerate}

\end{itemize}

\begin{figure*}[!th]
\subtable[Scenario A]{\includegraphics[scale=0.18]{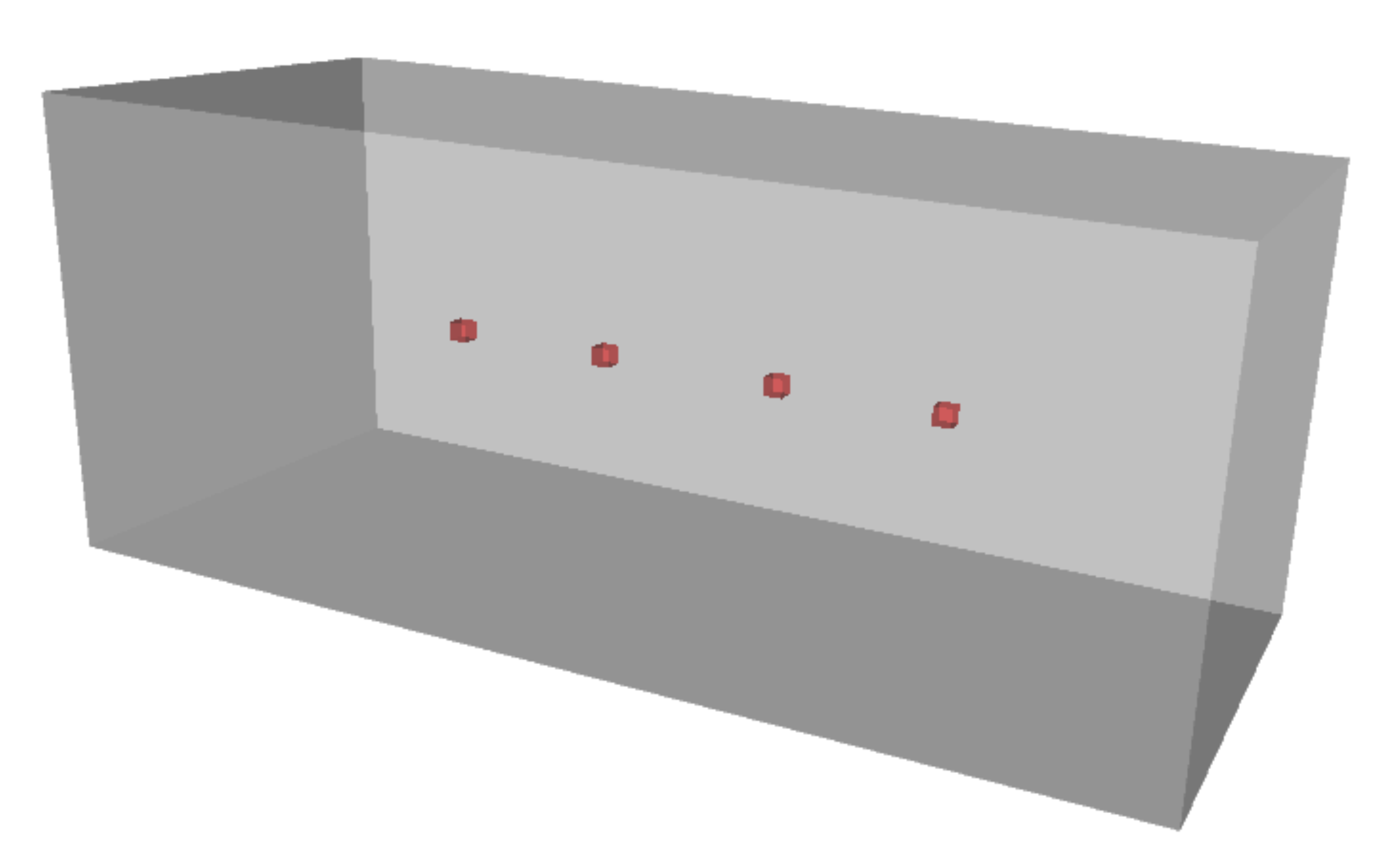}}%
\subtable[Scenario B]{\includegraphics[scale=0.18]{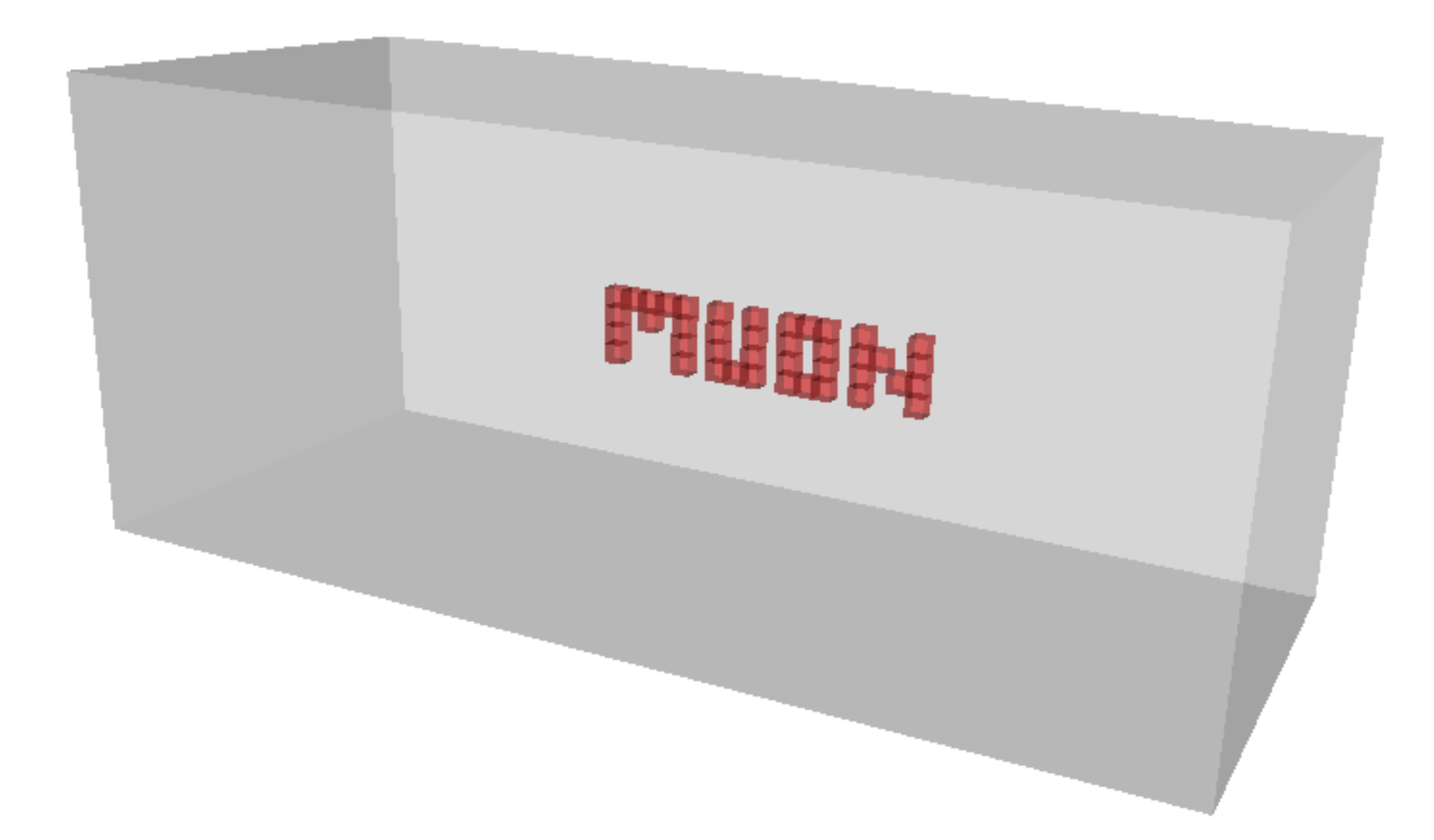}}%
\subtable[Scenario C]{\includegraphics[scale=0.18]{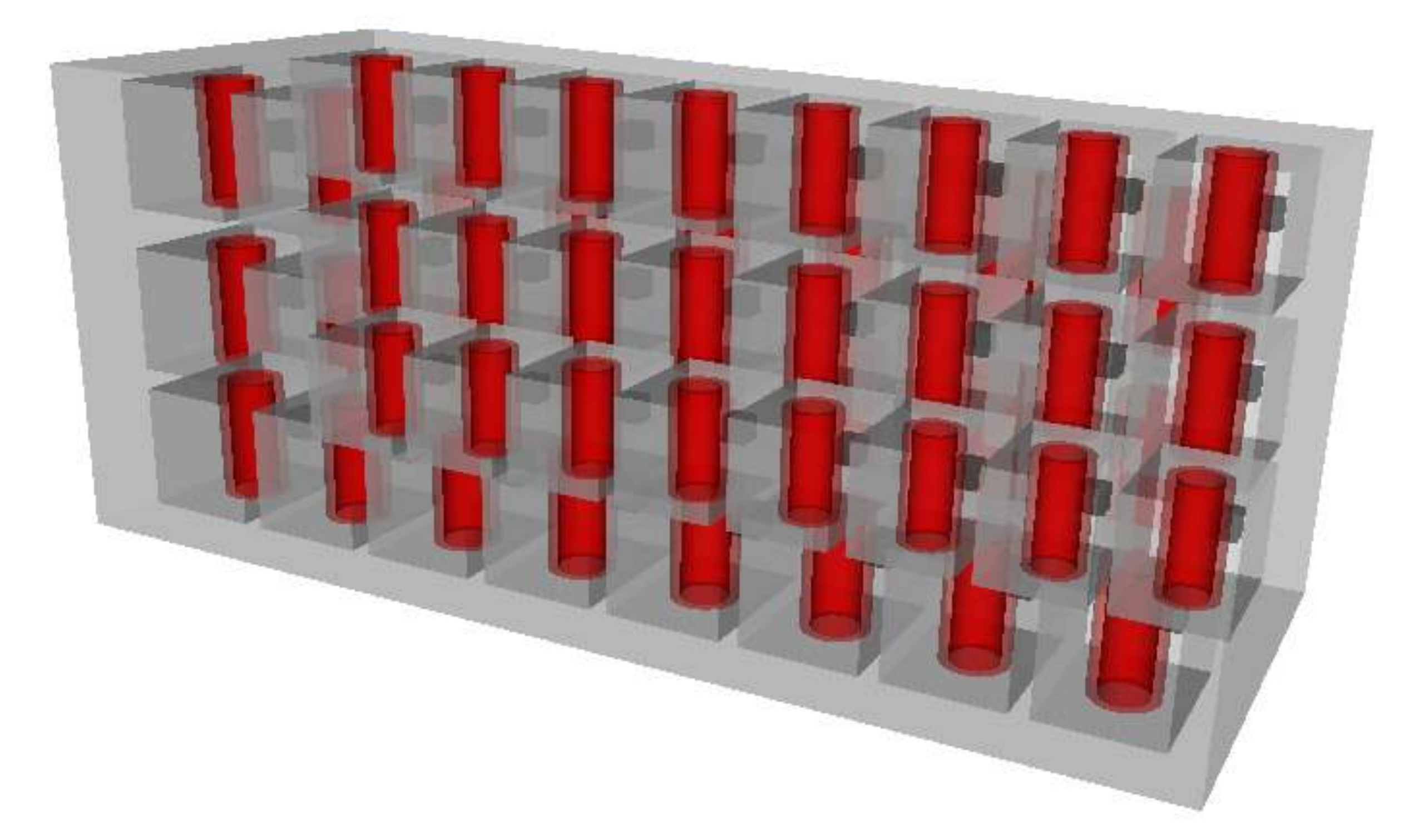}}%
\caption{Simulated tomographic scenarios used to validate the imaging algorithms.}%
\label{TomographyScenariosFig}
\end{figure*}

It is well known that with iterative algorithms the image reconstruction can be deteriorated as the iteration proceeds. An early stopping criterion is therefore
often used. Here we decided to stop the iterative procedure when the average relative $\lambda_{j}$ variation drops below a prespecified threshold 
$\varepsilon$, namely:
\begin{equation}
\frac{1}{N_{voxels}}\sum_{j}^{N_{voxels}} \frac{\lambda_{j}^{(k+1)}-\lambda_{j}^{(k)}}{\lambda_{j}^{(k)}}<\varepsilon\;\;\;\; 
\end{equation}
where we assumed $\varepsilon$=1\% and we required the criterion to be fulfilled during a given number of consecutive iterations (i.e. 5). 
Typically 20-30 iterations are needed to match the above criterion for the considered tomographic scenarios.

\begin{figure*}[!t]
\centering
\subtable[POCA XY tomography - Scenario A]{\includegraphics[scale=0.35]{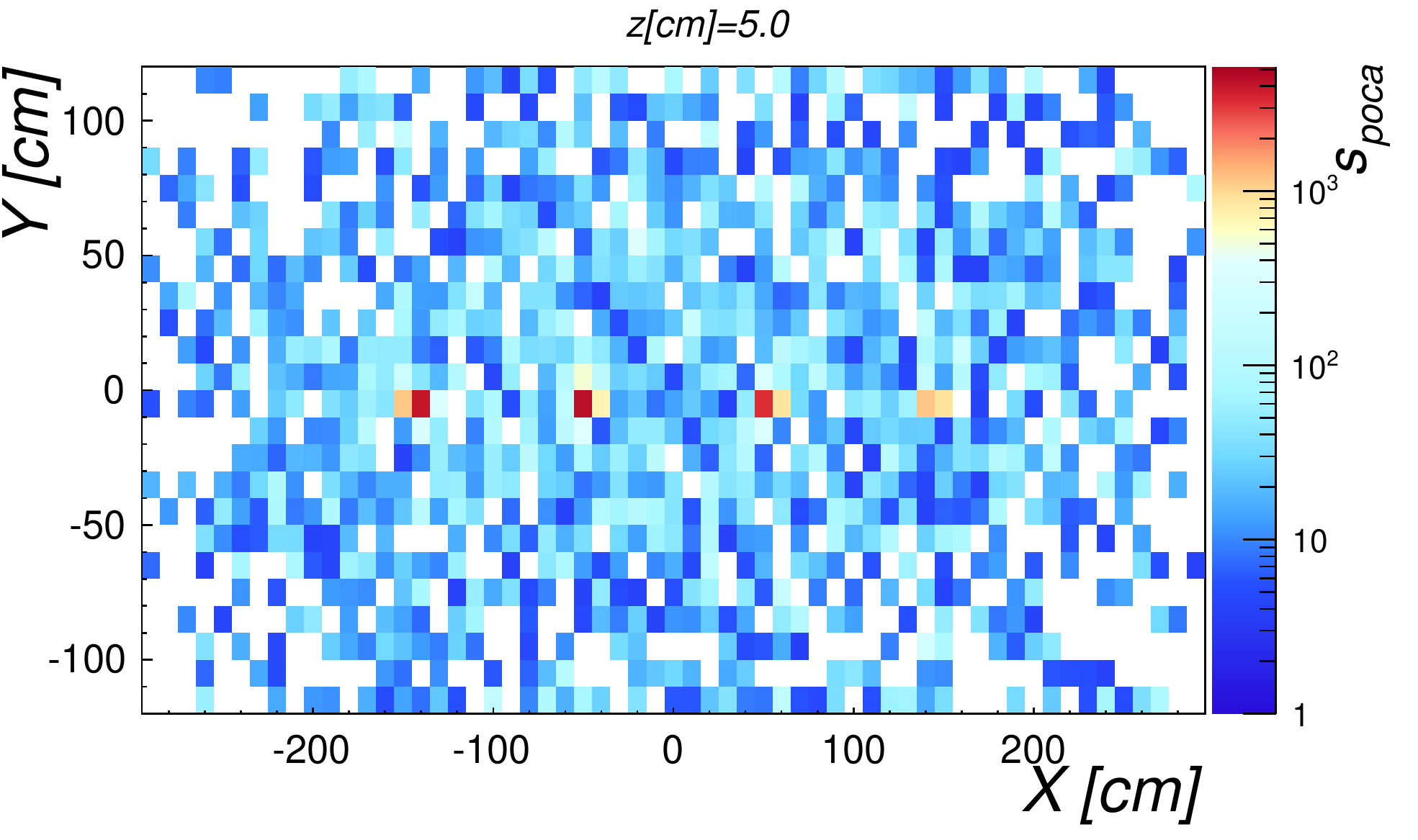}\label{POCAImagingFig1}}%
\hspace{0.15cm}
\subtable[POCA 3D rendering - Scenario A]{\includegraphics[scale=0.25]{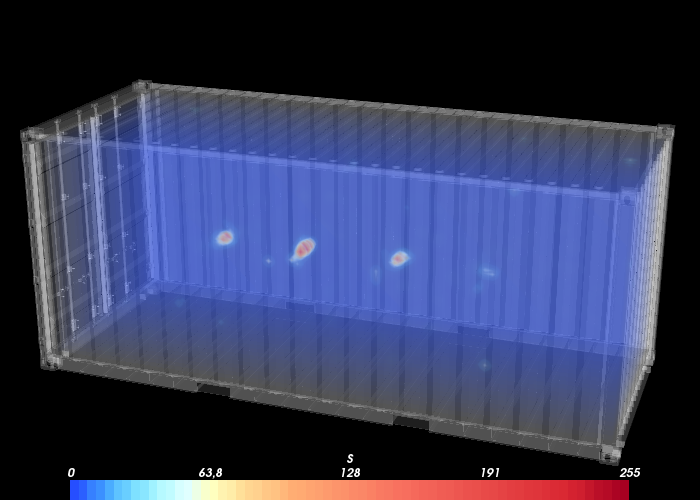}\label{POCAVolumeRenderingFig1}}\\%

\subtable[POCA XY tomography - Scenario B]{\includegraphics[scale=0.35]{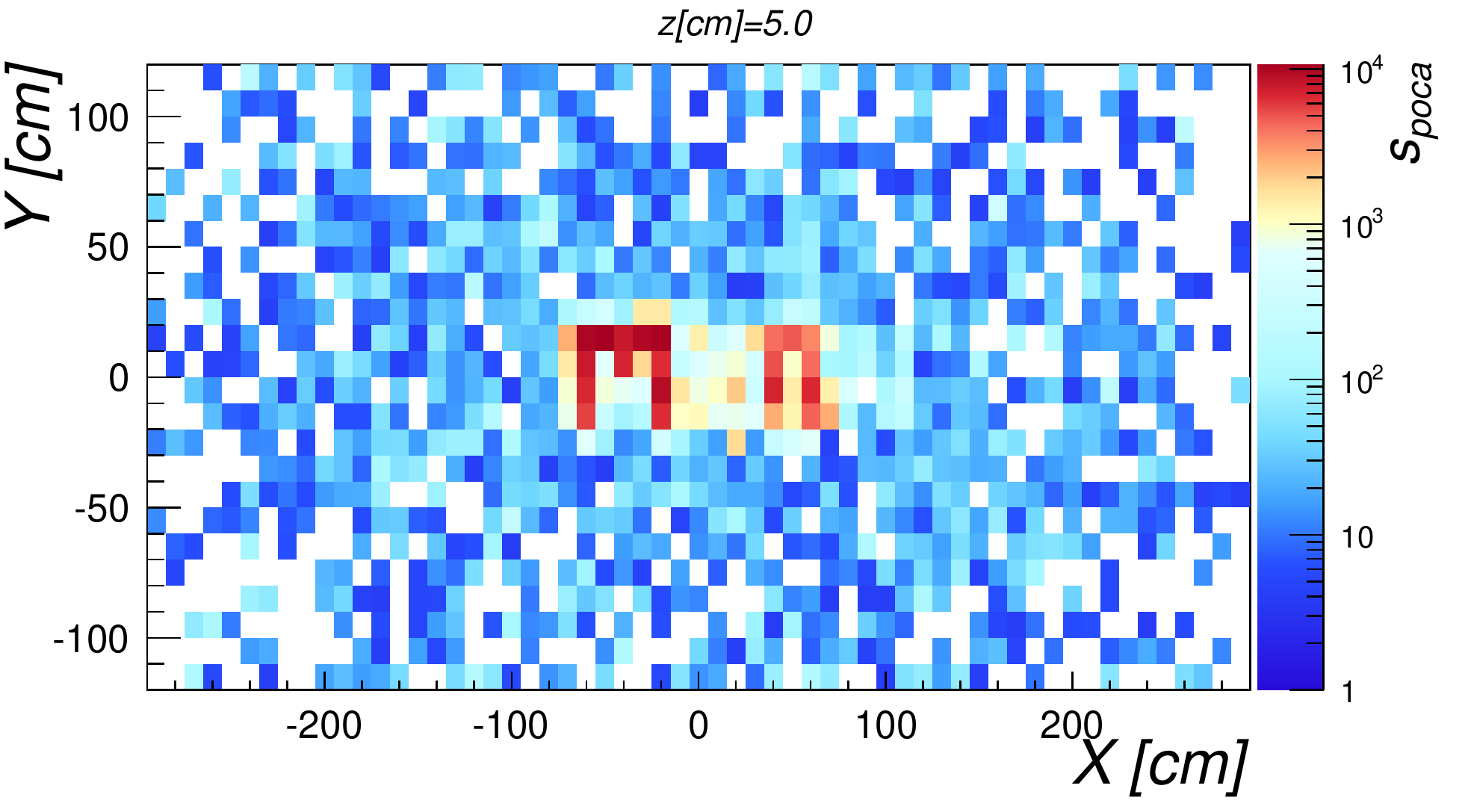}\label{POCAImagingFig2}}%
\hspace{0.15cm}
\subtable[POCA 3D rendering - Scenario B]{\includegraphics[scale=0.25]{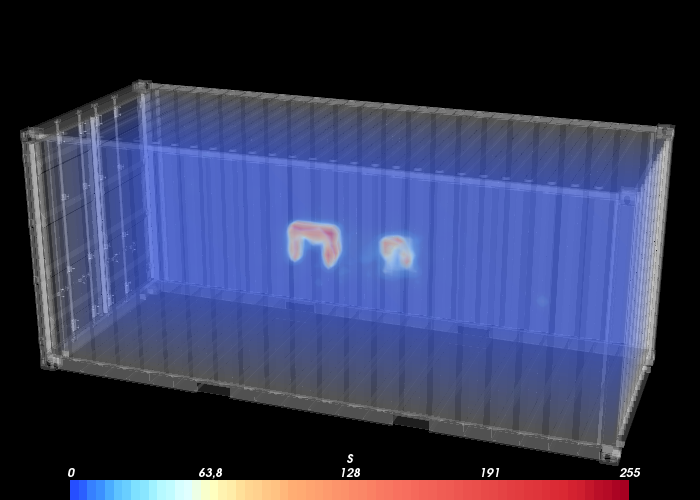}\label{POCAVolumeRenderingFig2}}\\%

\subtable[POCA XY tomography - Scenario C]{\includegraphics[scale=0.35]{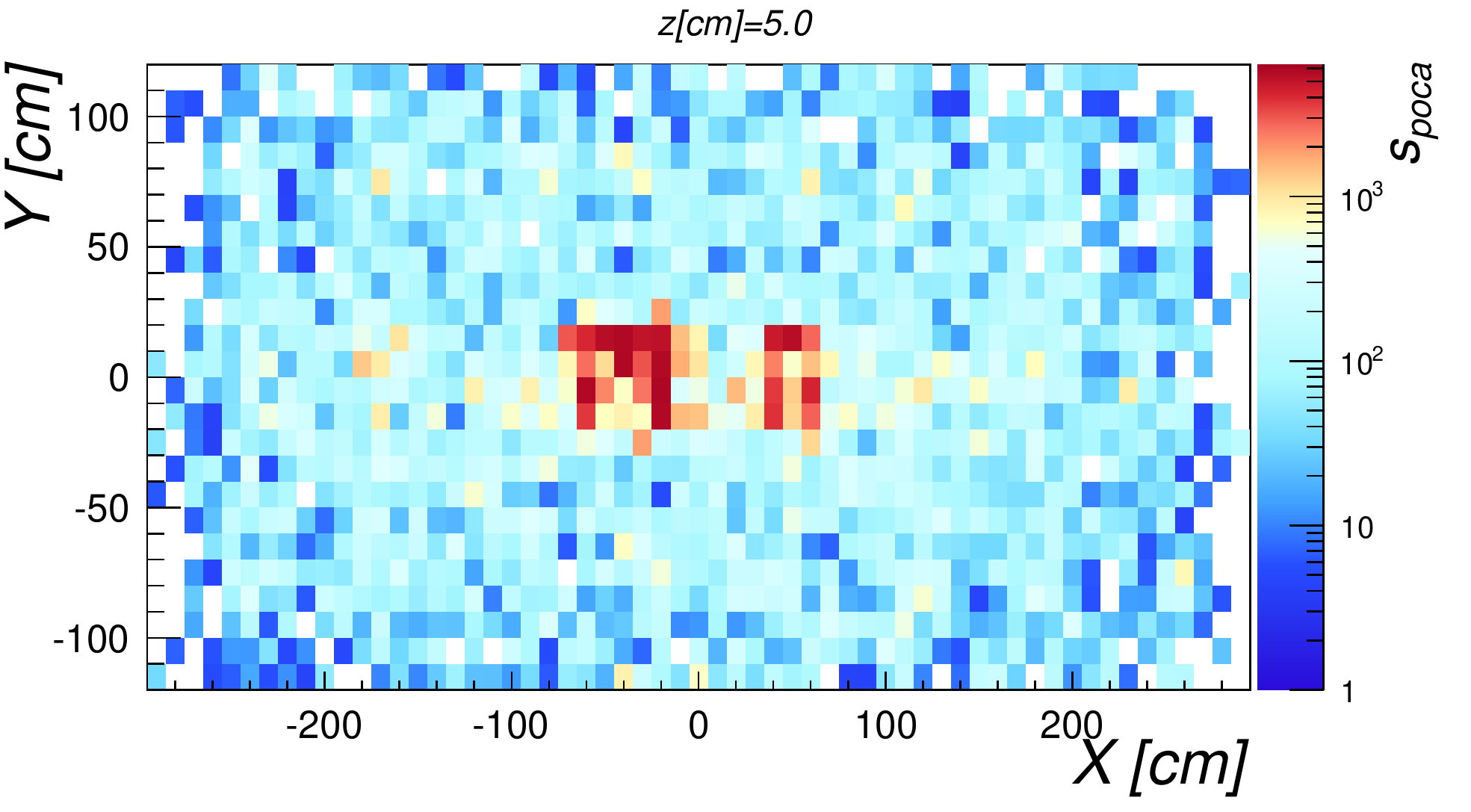}\label{POCAImagingFig3}}%
\hspace{0.15cm}
\subtable[POCA 3D rendering - Scenario C]{\includegraphics[scale=0.25]{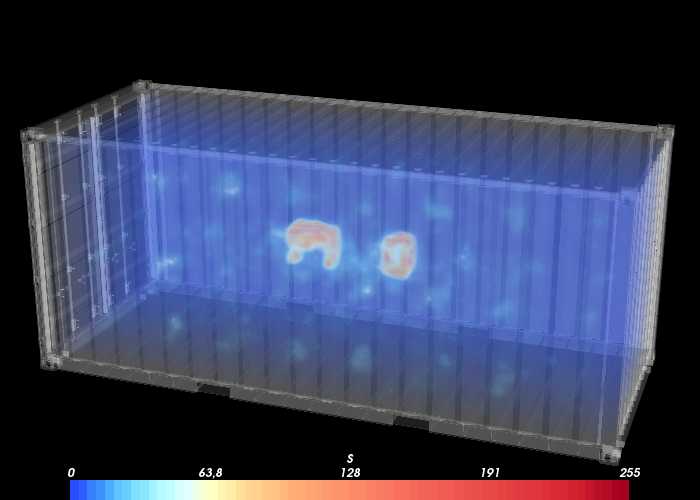}\label{POCAVolumeRenderingFig3}}%
\vspace{0.1cm}
\caption{Tomographic imaging of the three simulated scenarios (from top to bottom panels) obtained with the POCA method. In the left panels the XY tomography view
for a fixed $z$ depth ($z$= 5 cm) is shown while in the right panels a 3D volume rendering of the entire container is reported.}
\label{POCAFig}
\end{figure*}

\begin{figure*}[!t]
\centering
\subtable[2pt-ACF - Scenario A]{\includegraphics[scale=0.3]{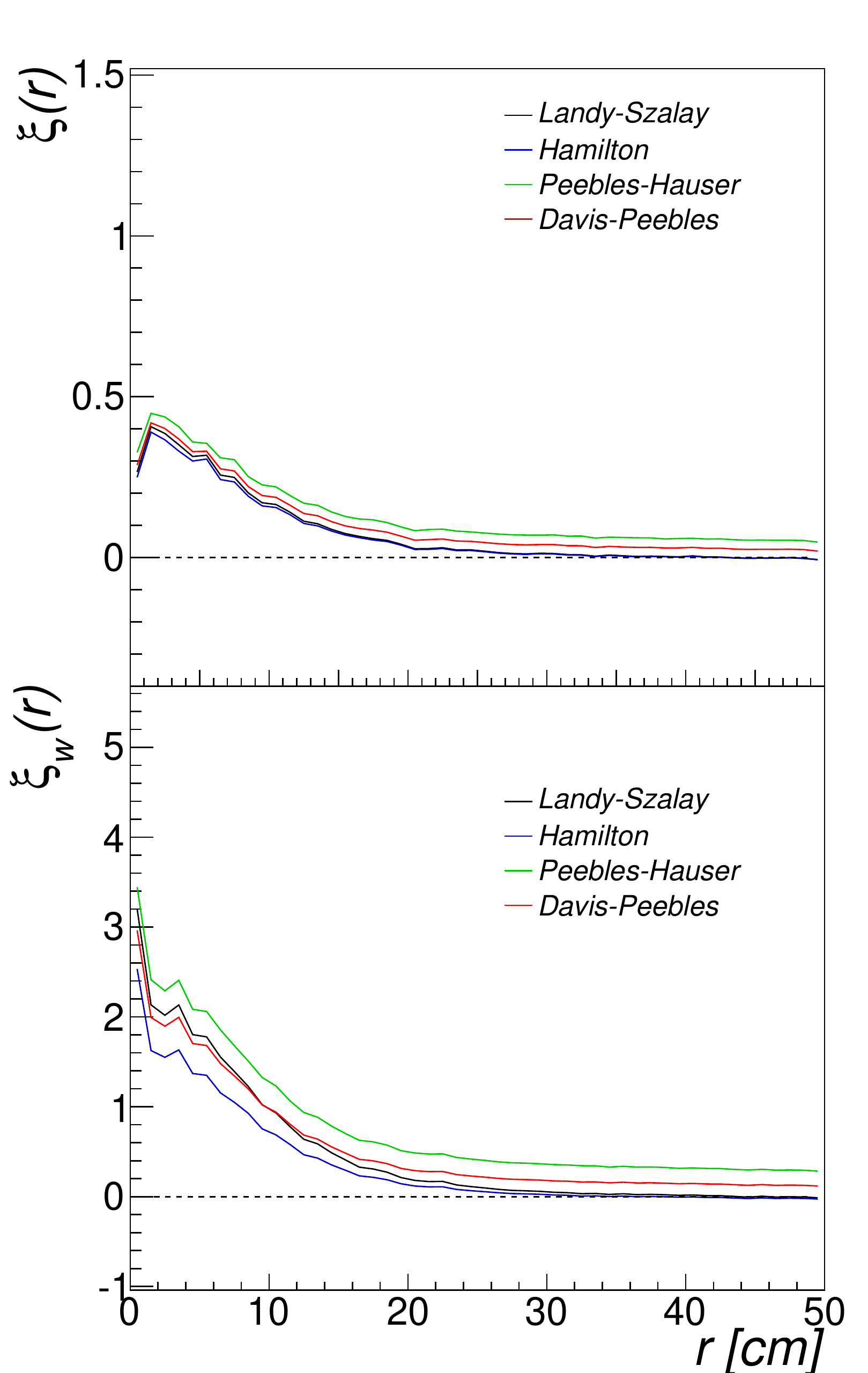}\label{ACFFig1}}%
\hspace{0.1cm}%
\subtable[2pt-ACF - Scenario B]{\includegraphics[scale=0.3]{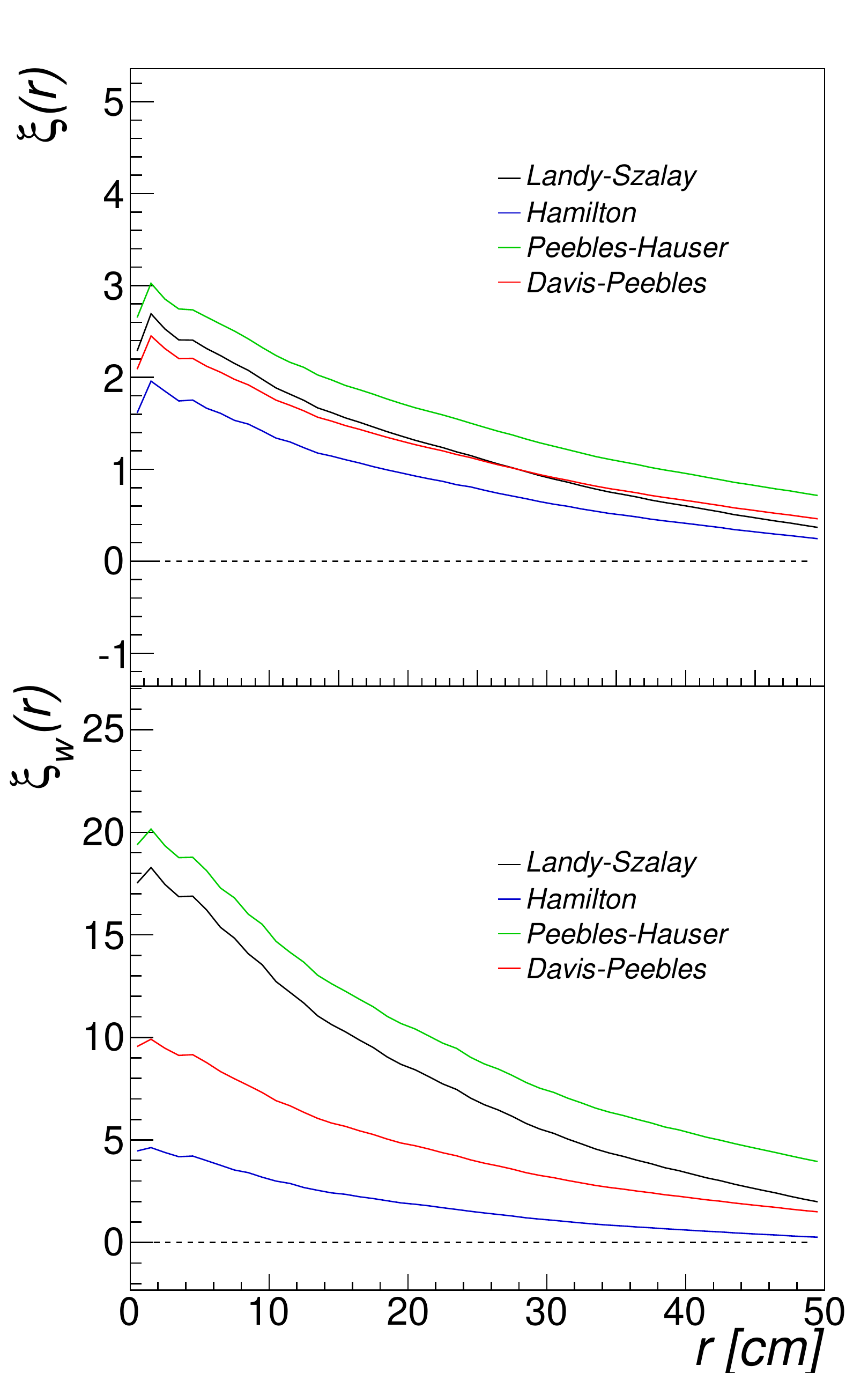}\label{ACFFig2}}%
\hspace{0.1cm}%
\subtable[2pt-ACF - Scenario C]{\includegraphics[scale=0.3]{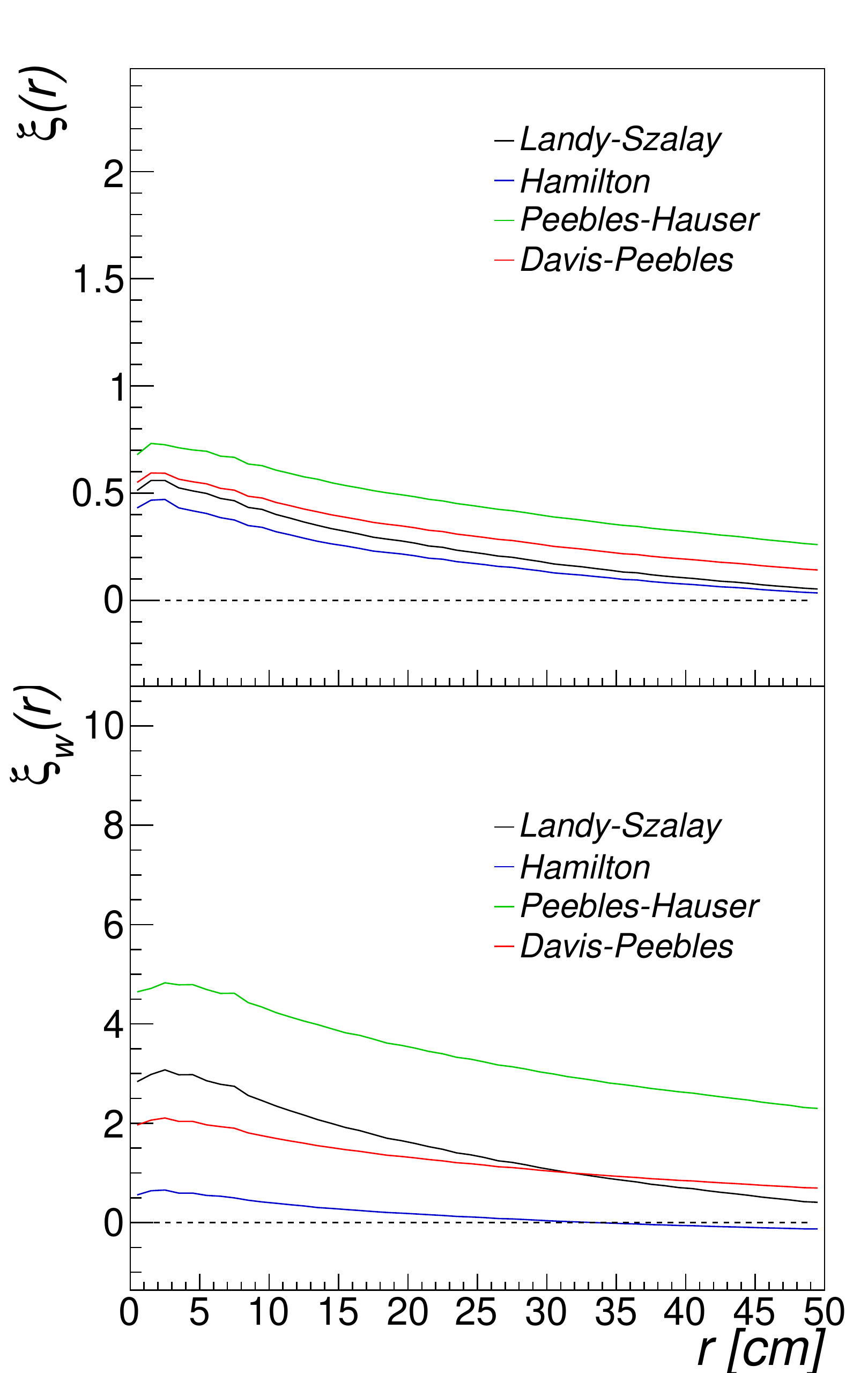}\label{ACFFig3}}%
\vspace{0.1cm}
\caption{2pt-ACF computed for the three tomographics scenarios from left to right panels. The colored lines refer to the four estimators described in 
the section \ref{AutocorrelationSection}.}
\label{ACFFig}
\end{figure*}

\section{Application to simulated data}\label{ResultsSection}
To validate the imaging methods described in the previous section in realistic conditions we developed a
detailed \textsc{Geant4} simulation of the detector. More details on the simulation procedure
as well as on the data reconstruction and quality selection are reported in the next sections (\ref{EventSimulationSection}, \ref{EventReconstructionSection}). 
Typical results obtained over different tomographic scenarios are presented in section \ref{ResultsSection}.

\begin{figure*}[!t]
\centering
\subtable[Clustering - Scenario A]{\includegraphics[scale=0.25]{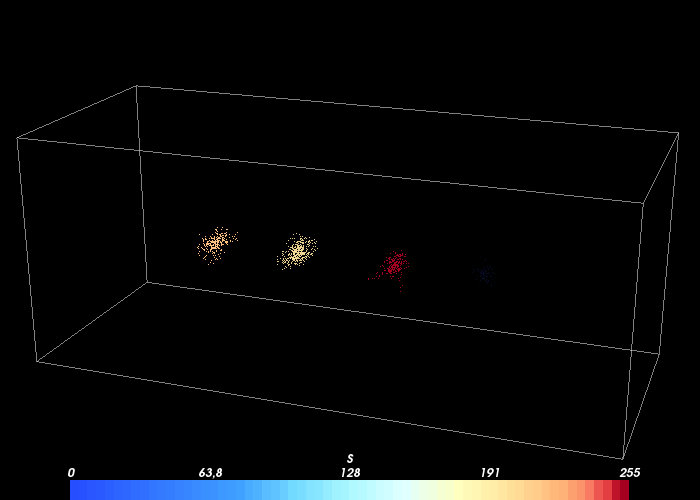}\label{ClusteringImagingFig1}}%
\hspace{0.15cm}
\subtable[Clustering 3D rendering - Scenario A]{\includegraphics[scale=0.25]{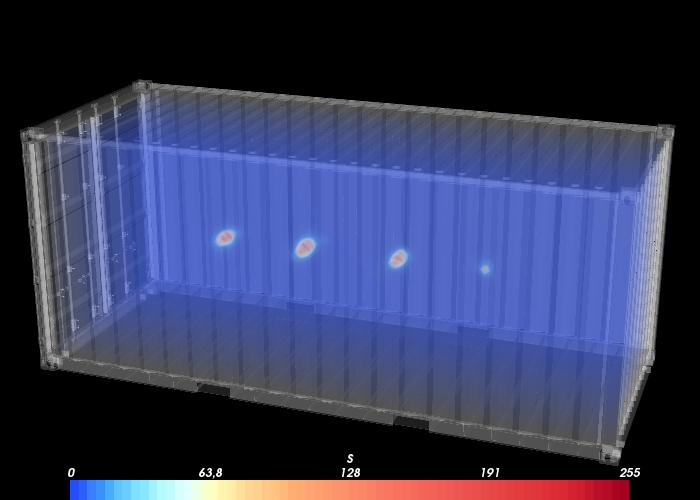}\label{ClusteringVolumeRenderingFig1}}\\%

\subtable[Clustering - Scenario B]{\includegraphics[scale=0.25]{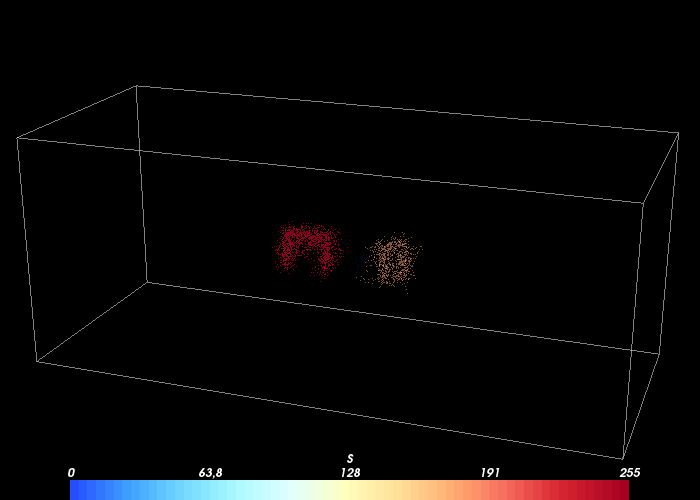}\label{ClusteringImagingFig2}}%
\hspace{0.15cm}
\subtable[Clustering rendering - Scenario B]{\includegraphics[scale=0.25]{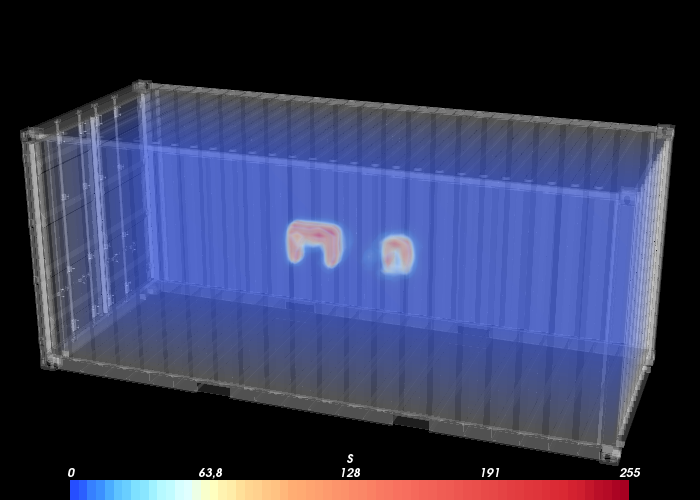}\label{ClusteringVolumeRenderingFig2}}\\%

\subtable[Clustering - Scenario C]{\includegraphics[scale=0.25]{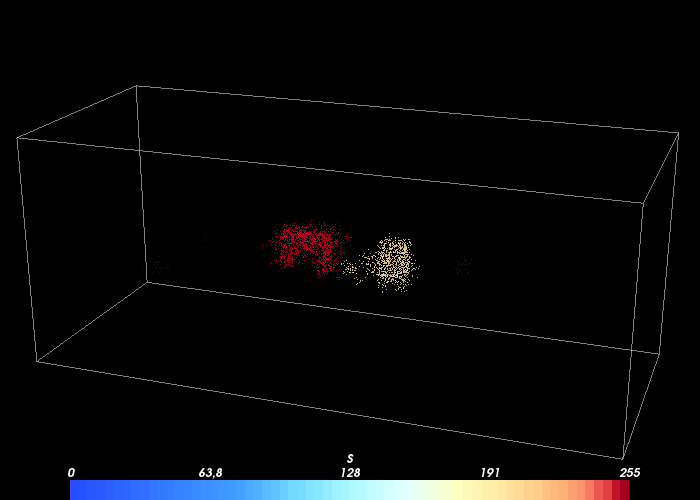}\label{ClusteringImagingFig3}}%
\hspace{0.15cm}
\subtable[Clustering 3D rendering - Scenario C]{\includegraphics[scale=0.25]{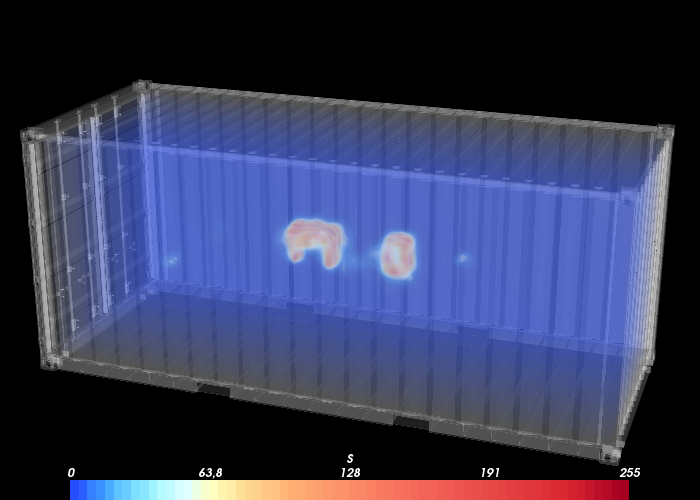}\label{ClusteringVolumeRenderingFig3}}%
\vspace{0.1cm}
\caption{Clustering results for the three tomographics scenarios from left to right panels.}
\label{ClusteringFig}
\end{figure*}

\subsection{End-to-end detector simulation}\label{EventSimulationSection}
The developed simulation incorporates all relevant detector elements (scintillators, WLS fibers, \dots) and also the relevant mechanical structures responsible
for the possible muon scatterings along its path. These includes the support rack, the container volume
(essentially roof and floor) eventually with potential threat objects and a ground layer placed below the detector structure.\\ 
Cosmic ray muons are injected in the detector with realistic energy and angular distributions, as derived from 
\textsc{Corsika} \cite{CORSIKA} simulations for proton-induced showers, generated for the Catania location 
(sea level, 37$^{\circ}$30'4"68 N, 15$^{\circ}$4'27"12 E, ($B_{x}$,$B_{z}$)= (27.16,-35.4)) with a E$^{-2.6}$ energy spectrum 
in the range 10$^{9}$-10$^{15}$ eV and isotropic angular distribution. The energy distributions of the secondaries is approximately log-normal peaked at $\sim$3 GeV. 
The angular distributions are $\propto\sin\theta\cos^{2}\theta$, peaked at $\sim$30$^{\circ}$.
The particle distributions obtained with \textsc{Corsika} have been parametrized for fast generation.\\
Two kind of simulations are available. Full simulations include the explicit trasport of photons inside the scintillator bars and WLS fibres and are typically
used only for detector design studies. Fast simulations, in which optical processes are switched off, are instead used for event reconstruction studies as well
as to provide tomographic scenarios to test the imaging algorithms.\\In Figure \ref{DetectorFig} we show a sample simulated 
$\mu$ event of energy 1 GeV together with hits produced in each detector plane. The color scale in the plots represents the
energy deposited in each hit. Hits with red borders are effectively due to the muon while others are spurious.

\subsection{Event reconstruction and selection}\label{EventReconstructionSection}
The event reconstruction procedure is done according to the following stages:
\begin{itemize}
 \item \textit{Hit selection}: Strips are considered to be triggered if the particle energy deposit $dE/dX$ (or the number of produced photoelectrons) is 
 above a pre-specified threshold. A trigger threshold of 1 MeV is assumed for strip triggering. Furthermore we select strips triggering within a 
 given time interval $\Delta t$. Events triggering at least four planes, hereafter denoted as 4-fold events, are selected and hits from each X-Y planes 
 are collected to form a list of candidate track points. Each of these points
 is smeared with a Gaussian of width equal to the detector position resolution $\sigma$=1 cm/$\sqrt{12}\sim$2.9 mm.
 \item \textit{Cluster finding}: The list of candidate track points in each plane is scanned to find cluster candidates, defined by adjacent hit strips. The obtained 
 clusters can be eventually re-splitted in single hits afterwards if the cluster multiplicity is above a given threshold. Finally single hits are replaced with the cluster
 barycenter and passed to the track finding stage.
 \item \textit{Track finding}: Valid track candidates (at least one hit in each plane, $\theta_{rel}<60^{\circ}$) are collected and then reconstructed 
 using a Kalman-Filter approach \cite{KalmanFilter}. In case of multiple track candidates, the track selection is done according to minimum $\chi^{2}$ criterion.
\end{itemize}
For tomography studies a practical choice is to select events with only one cluster per plane. Multi-cluster events with larger multiplicity will be 
anyway recorded for cosmic ray physics studies.\\
To reject spurious events and therefore to reduce the chance of getting false positive in the imaging phase, we applied in the reconstruction algorithms 
a further quality selection to the available data:
\begin{itemize}
 \item POCA, Clustering, 2pt-ACF: Events with scattering angles $\theta$ larger than 2 degrees are selected to reduce the noise due to other scatterers;
 \item EM-ML: Only events crossing the entire container from the top plane to the bottom plane are considered.
 To limit the chances of uncorrect raytracing, leading to misidentifications and fakes, the POCA information is used to define the muon path 
 inside the container only for events with a ``trustable'' POCA reconstruction, e.g. those preliminarly selected  
 in the clustering analysis stage.
 After the selection chain, only a small percentage ($\sim$5\%) of the total events was found to be rejected.
\end{itemize}

\begin{figure*}[!t]
\centering
\subtable[EM-ML XY tomography - Scenario A]{\includegraphics[scale=0.3]{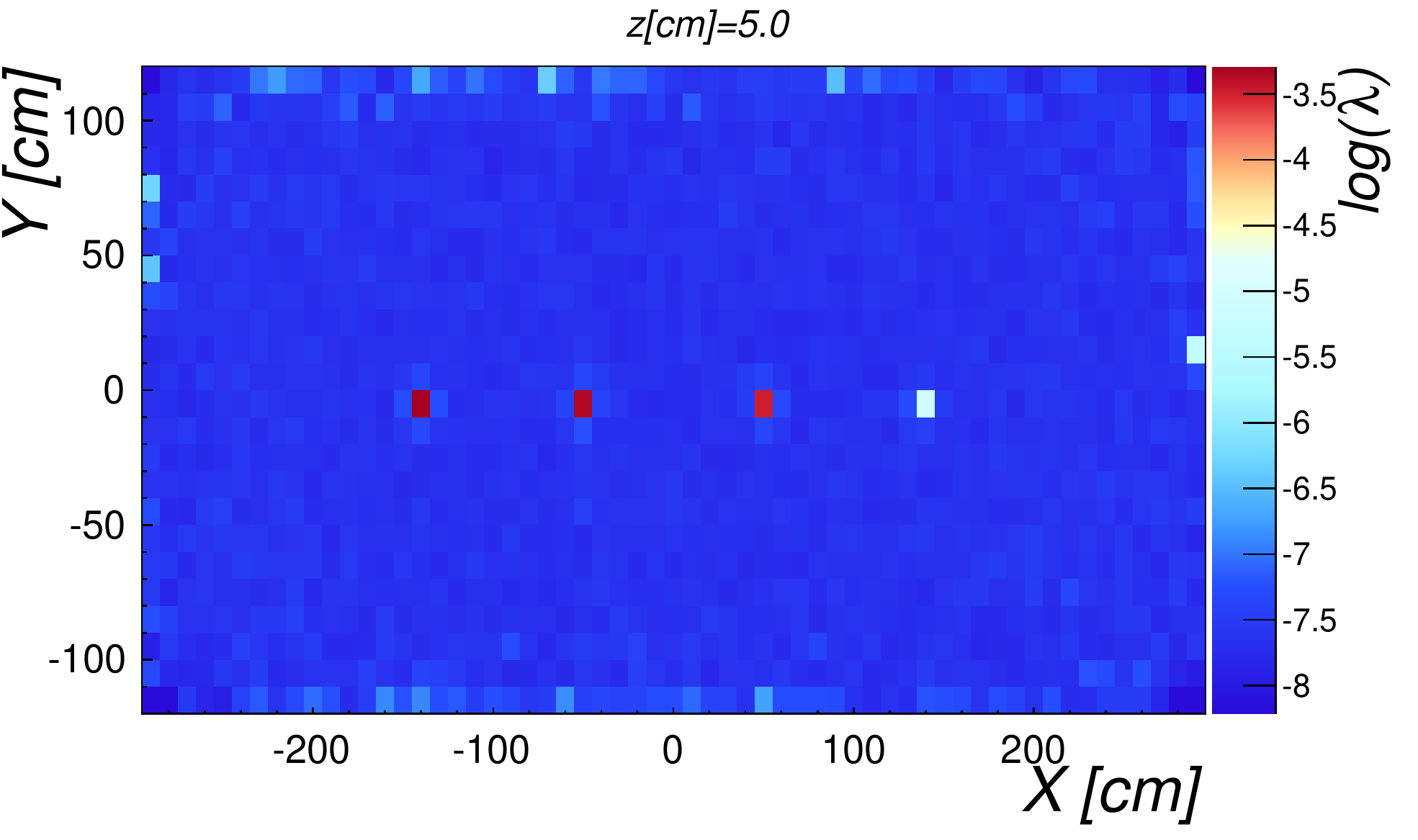}\label{EMLLFig1}}%
\hspace{0.15cm}
\subtable[EM-ML 3D rendering - Scenario A]{\includegraphics[scale=0.25]{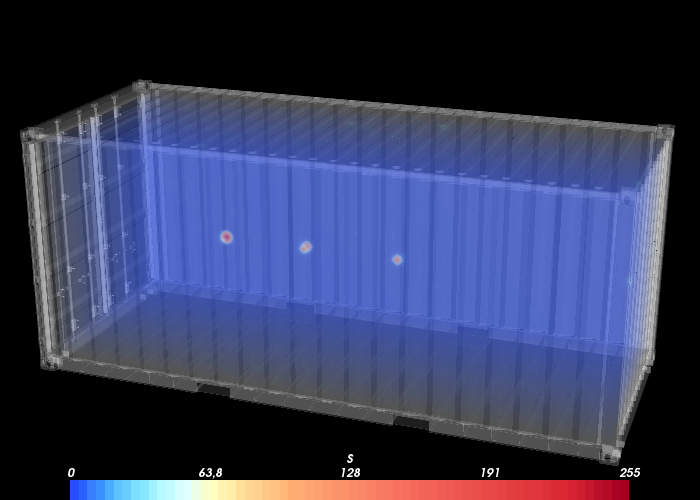}\label{EMLLFig2}}\\%

\subtable[EM-ML XY tomography - Scenario B]{\includegraphics[scale=0.3]{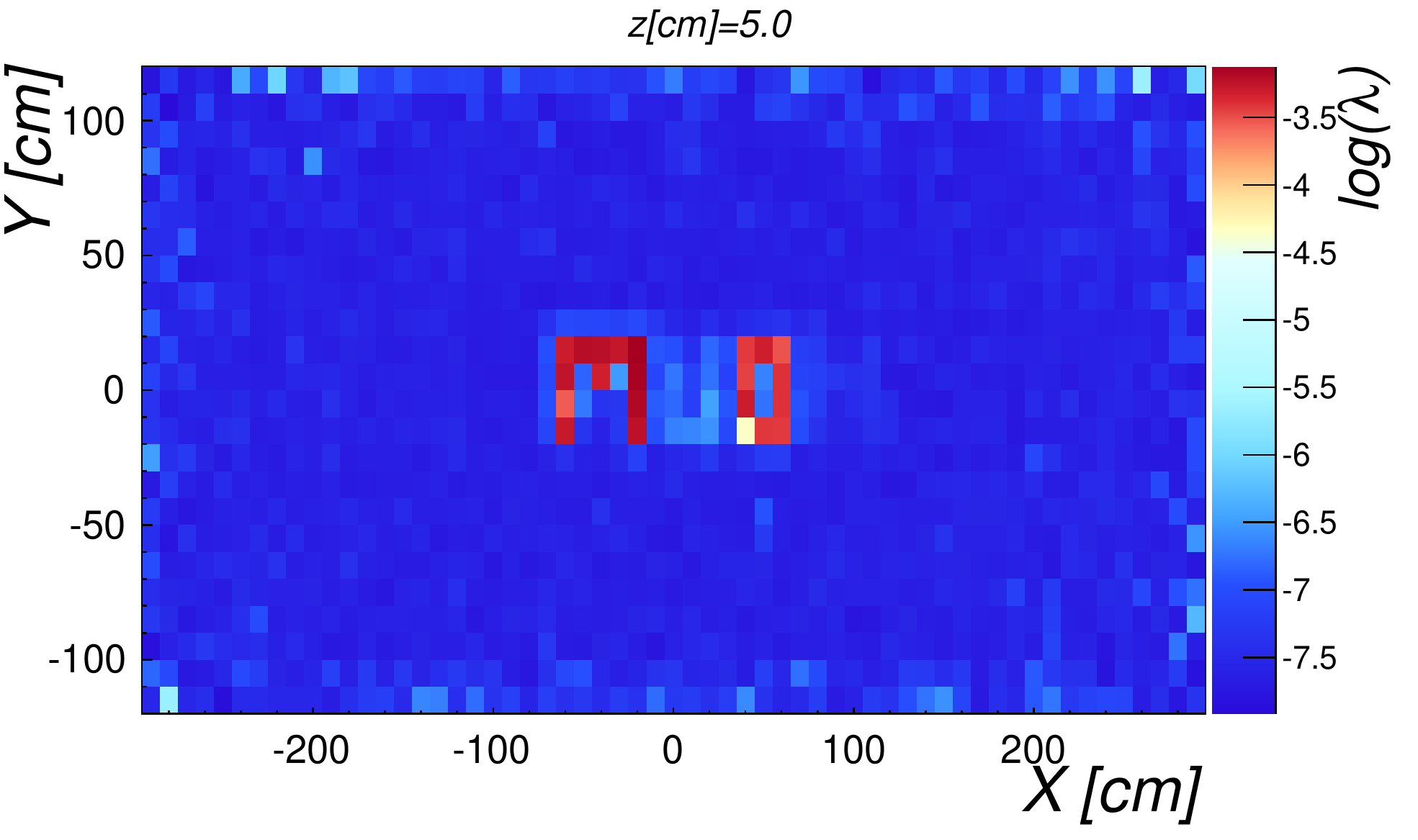}\label{EMLLFig3}}%
\hspace{0.15cm}
\subtable[EM-ML 3D rendering - Scenario B]{\includegraphics[scale=0.25]{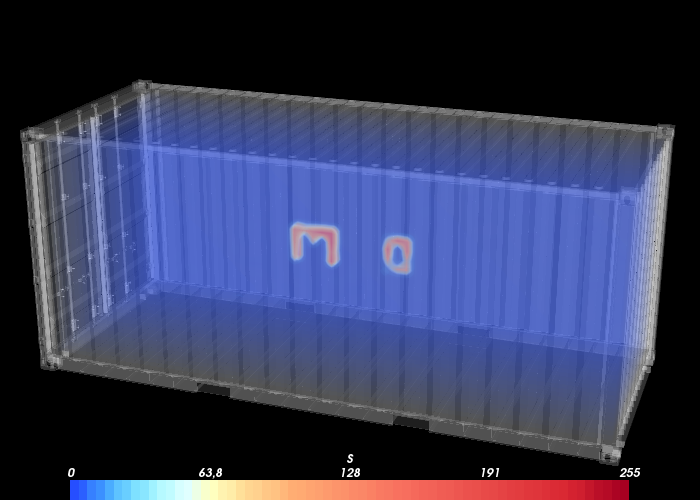}\label{EMLLFig4}}%

\subtable[EM-ML XY tomography - Scenario C]{\includegraphics[scale=0.3]{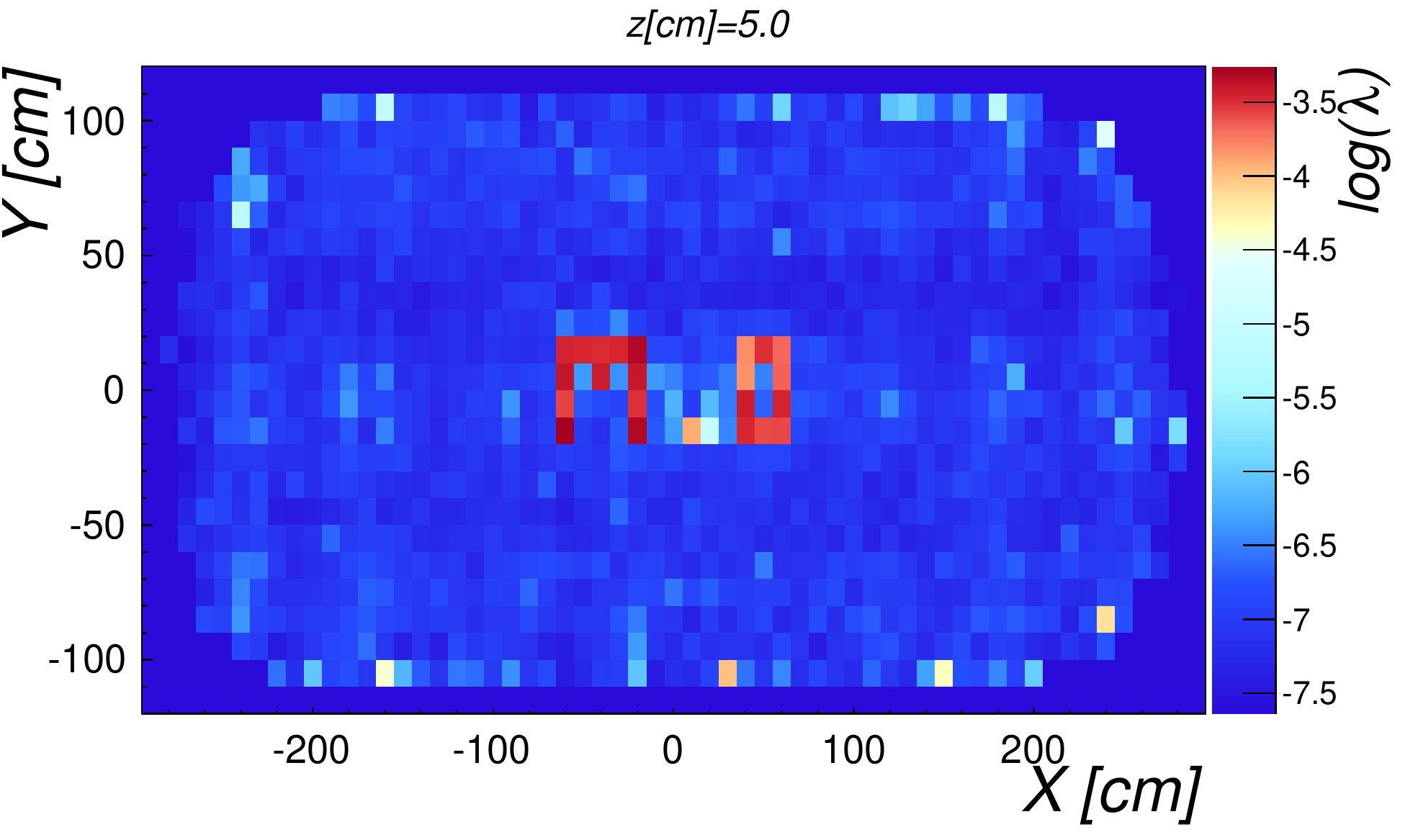}\label{EMLLFig5}}%
\hspace{0.15cm}
\subtable[EM-ML 3D rendering - Scenario C]{\includegraphics[scale=0.25]{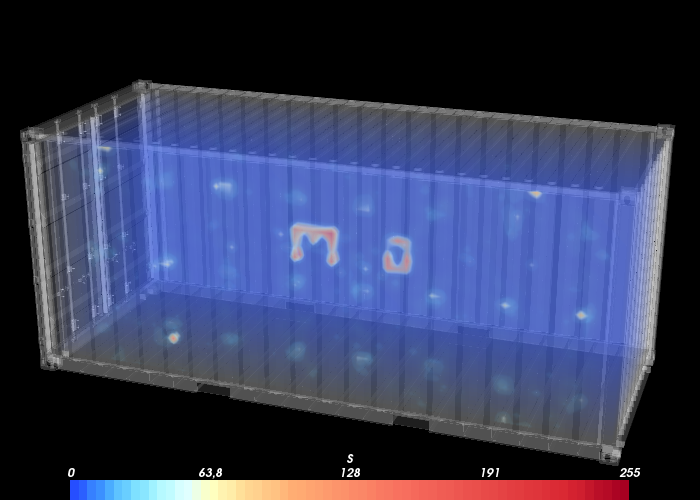}\label{EMLLFig6}}%
\vspace{0.1cm}
\caption{Tomographic imaging of the three simulated scenarios (from top to bottom panels) obtained with the EM-ML method. In the left panels the XY tomography view
for a fixed $z$ depth ($z$= 5 cm) is shown while in the right panels a 3D volume rendering of the entire container is reported.}
\label{EMLLFig}
\end{figure*}

\subsection{Imaging results}\label{ImagingResultsSection}
We validated the implemented algorithms using \textsc{Geant4} simulations with different tomographic scenarios, shown in Figure \ref{TomographyScenariosFig}:
\begin{itemize}
 \item \emph{Scenario A}: Four threat boxes ($W$, $U$, $Pb$, $Sn$) of size 10 cm$\times$10 cm$\times$10 cm inserted 
 at the center of a empty container. The container load relative to the scene is $\sim$100 kg.
 \item \emph{Scenario B}: A ``\textsc{muon}'' shape built with voxels of size 10 cm$\times$10 cm$\times$10 cm inserted 
 at the center of an empty container. Each letter is made of different materials: \textsc{m}= Uranium, \textsc{u}= Iron, \textsc{o}= Lead,
 \textsc{n}= Aluminium. The container load relative to the scene is $\sim$480 kg.
 \item \emph{Scenario C}: Same of scenario B. A denser environment is assumed inside the container volume, filled with layers of washing machine-like elements. 
 These are made by an aluminium casing with an iron engine inside with relative support bars and a concrete block. The container load relative to the scene 
 is $\sim$3500 kg.
  
\end{itemize}
A number of muon events of 5$\times$10$^{5}$, corresponding to $\sim$10 minutes scanning time, has been simulated for each scenario using a realistic energy spectrum with 
range 0.1-100 GeV.\\In Figure \ref{POCAFig} we report the results relative to the POCA method for the three scenarios under test. On the left panels we report the
tomographic XY section of the container for a fixed $z$ level equal to $z$= 5cm (volume center). The right panels show a 3D volume rendering of the entire container. 
The volume has been divided into cells of volume 10 cm$\times$10 cm$\times$10 cm and the color scale represents the POCA signal for each bin $i$, namely 
$\sum_{j=1}^{N_{i}}\theta_{j}^{2}$ with $N_{i}$ number of POCA events in bin $i$. As can be seen, all three scenarios are successfully identified. Due
to the intrinsic resolution of the POCA method\footnote{We performed dedicated simulations of cosmic ray muons traversing a single uranium layer of thickness 
10 cm and reconstructed the POCA information for each event. About 20\% of the events have the POCA information reconstructed outside the expected threat volume, 
falling in particular in the first surrounding voxels.} a persistent halo is present and consequentely the imaged objects are slightly increased in size 
with respect to the real dimensions, particularly along the vertical z-axis. A considerable noise, related to the engine elements, is present in the dense 
environment scenario. In such case we needed to adopt a more stringent quality cut in the scattering angle ($\theta>$6$^{\circ}$), to achieve the identification of 
the threat objects.\\We report in Figure
\ref{ACFFig} the results relative to the autocorrelation analysis, computed using a random data sample of 5$\times10^{6}$ simulated events in an empty container volume,
e.g. ten times larger than the data sample under investigation. Upper plots refer to the standard 2pt-ACF estimators, reported with different color lines, 
while in the bottom panels we report the weighted correlation function. As can be seen, in all cases we obtain a significant excess with respect to the background at
distance scales around 5 cm. The excess can be largely enhanced by using the scattering angle information (weighted 2pt-ACF) together with the spatial information. 
\\In Figure \ref{ClusteringFig} we
report the results obtained with the clustering method for the three scenarios. The correlation scale found with ACF analysis provides a useful estimation of 
the linking length parameter $\epsilon$ to be used in the clustering reconstruction. We assumed $\epsilon$=5 cm and a minimum point threshold
of $N_{pts}$= 30. The color scale for the $i$-th cluster indicates the cluster weight $w_{i}$, defined as 
$w_{i}$= $\sum_{j=1}^{N_{i}}\theta_{j}^{2}/N_{j}$ with $N_{i}$ number of POCA points in cluster $i$. As one can see, a good accuracy is achieved in the identification
of the three different scenarios, with a larger presence of noisy clusters in the reconstructed scenario C.\\Finally in Figure \ref{EMLLFig} we report the results obtained
with the EM-ML method. As can be observed the target objects are reconstructed with a considerably better resolution. The halo responsible for the object deformation 
observed with the POCA reconstruction is almost absent. As already discussed for the other algorithms, the last scenario
required a different strategy with respect to the scenarios A and B. In particular we observed that in presence of noise the convergence criterion adopted for the other 
reconstructions ($\epsilon$=1\%) is not sufficient to achieve a tomography image with quality comparable to scenario B. A smaller tolerance parameter was therefore assumed 
($\epsilon=$0.05\%) to achieve better results (see Figs. \ref{EMLLFig5}, \ref{EMLLFig6}), at the cost of significantly increasing the number of required iterations and
the computing times. This scenario demonstrates that a finer tuning of the likelihood algorithm is currently desirable over different, more realistic, noisy scenarios 
with the aim of determining a unique configuration for real time analysis.\\To assess the reconstructed image quality we made use of the structural similarity index 
\emph{SSIM} introduced in \cite{SSIMPaper} for two-dimensional images, extended to our three-dimensional images. 
It allows luminosity, contrast and structure comparisons on a local basis between the reconstructed tomographic image and 
a reference image, built by considering the known scattering densities in each scenario as defined in formula \ref{ScatteringDensityDefinition}.
After proper normalization of both maps to the same range, we considered a comparison window around each image pixel $j$ and calculated over it the pixel means 
($\mu_{j,rec}$, $\mu_{j,ref}$), standard deviations ($\sigma_{j,rec}$, $\sigma_{j,ref}$) and covariance $\sigma_{j,rec-ref}$ for the reconstructed 
and reference images respectively. The \emph{SSIM} index for pixel $j$ is then defined as:\\
\begin{align}
\mbox{\emph{SSIM}}_{j} = \frac{(2\mu_{j,rec}\mu_{j,ref}+C_{1})(2\sigma_{j,rec-ref}+C_{2})}{(\mu_{j,rec}^{2}+\mu_{j,ref}^{2}+C_{1})
(\sigma_{j,rec}^{2}+\sigma_{j,ref}^{2}+C_{2})}
\end{align}
where $C_{1}$ and $C_{2}$ are small constants introduced to avoid numerical instabilities when $(\mu_{j,rec}^{2}+\mu_{j,ref}^{2})$ or 
$(\sigma_{j,rec}^{2}+\sigma_{j,ref}^{2})$ are very close to zero. An index close to one indicates strong agreement with the reference image, while,
on the contrary, a nearly null index is symptomatic of a bad reconstruction.
It is possible to define also a mean similarity
index \emph{MSSIM}, obtained by averaging the previous index over all pixels or over a given region of interest.\\In Table \ref{SSIMIndexTable} we report the mean 
similarity index computed for the three tomographic scenarios under study and for the POCA, clustering and EM-ML methods. We considered a 3$\times$3$\times$3 
pixel window around each pivot pixel and we calculated the mean similarity index over a spatial region surrounding the threats. 
As the visual analysis already suggested, the computed indexes are close to unity, indicating an overall accurate reconstruction. The computed index effectively 
reflects the better reconstruction performances of the EM-ML method with respect to the other implemented methods.

\begin{table}[!th]
\centering%
\footnotesize%

\begin{tabular}{c|c c c|}
\hline%
\textbf{Scenario} & \multicolumn{3}{>{}c|}{\textbf{MSSIM}}\\%
& \emph{POCA} & \emph{Clustering} & \emph{EM-ML}\\%
\hline%
A & 0.94 & 0.94 & 0.99\\ %
B & 0.89 & 0.89 & 0.98\\ %
C & 0.83 & 0.83 & 0.89\\%
\hline%
\end{tabular}
\caption{Similarity index SSIM obtained with the POCA, clustering and EM-ML methods for the three tomographic scenarios under analysis.}
\label{SSIMIndexTable}
\end{table}

\section{Towards a real-time tomographic analysis}\label{SummarySection}

Concerning the software tools, two working lines are currently under progress within the project in view of the complete operation of the Muon Portal, one
aiming to develop a graphical user interface for the tomography and visualization tasks, and the other focusing on the optimization of the designed 
algorithms for real-time application.\\
To be compatible with the real container traffic at the harbours, the tomography analysis must be performed in reasonable small times, few minutes at most.
The POCA algorithm, with its simplicity, guarantees the smallest computation times, perfectly matching the port requirements. No optimizations are therefore 
needed in such case. This is not the case for the other designed algorithms.\\
The EM-ML algorithm typically requires $\sim$30 minutes on a Xeon QuadCore E5620 2.40Ghz processor for a typical scanning run of $\sim$ 10 minutes and 20-30 iterations, 
and therefore cannot match the requirements of a real time image processing, at least in its serial implementation. However, both the init and imaging 
step of the algorithm are
embarrassingly parallelizable as being based on independent event loops. We are therefore planning to implement also a parallel version of the algorithm using the 
\textsc{MPI} library \cite{MPI}. The achieved speed-up with respect to a serial implementation would be remarkable. 
The parallel implementation allows a real time application of the method even with a modest number of computing machines.\\
The computation of the 2pt ACF is a very time-consuming task, proportional to $N^{2}$ ($N$ size of the data sample).
Optimized serial implementations, based on building a kd-tree \cite{KdTree} with the data, allow to drop the algorithm complexity at the level of $N\log(N)$. Significative 
speed-up can then be obtained afterwards with ad hoc optimization and parallelization techniques \cite{ACFParallelized} or by making use of GPUs \cite{ACFGPU}.
At the present status a brute implementation is available. To maintain the
computation time at reasonable level, the pair calculation is limited to adjacent three-dimensional voxels with size matching the maximum desired correlation scale
and to observations with scattering angles larger than a predefined threshold. An optimized version of the algorithm is however currently being designed.\\
The density-based clustering algorithm suffers from the same problematic discussed for the ACF computation, as it requires the computation of the 
nearest neighbour of each point in the volume (O(N$^{2}$)). The algorithm has been optimized by using kd trees and further optimization strategies to 
achieve a O($N\log(N)$) complexity which makes the algorithm reliable for real time analysis.\\In conclusion we are pursuing a large efforts in combining different 
reconstruction and visualization tools for a reliable and fast image processing of a muon tomography. While standard algorithms have already been implemented
and their use in a real time processing of a tomographic image may be achieved even by a single standard processor, the use of alternative, more accurate, algorithms
requires additional work, possible on their parallelization, to provide a comparatively fast tool. Work along this line has already started and the results will
be reported in a future paper.

\end{document}